\documentclass[preprint,review,12pt]{elsarticle}
\usepackage[T1]{fontenc}  
\usepackage[scaled=0.82]{beramono}  
\usepackage{microtype} 
\usepackage{amsmath}
\usepackage{subcaption}
\usepackage{float}
\usepackage{amsfonts}
\usepackage{mathtools}
\usepackage{bm}
\usepackage{slashed}
\usepackage{amssymb}
\usepackage{hyperref}
\usepackage{lineno}
\hypersetup{
    colorlinks=true,
    linkcolor=black,
    citecolor=black,
    urlcolor=black
}
\usepackage{nameref}
\usepackage{siunitx}
\usepackage{mhchem}
\usepackage{subcaption}
\usepackage[margin=2cm]{geometry}
\journal{Elsevier}

\begin{document}
\begin{frontmatter}
    \title{{Single electron charge spectra of 8-inch  high-collection-efficiency MCP-PMTs}}
    \author[a,b,c]{Jun Weng}
    \author[a,b,c]{Aiqiang Zhang}
    \author[d]{Qi Wu}
    \author[d]{Lishuang Ma}
    \author[a,b,c]{Benda Xu\corref{cor1}}%
    \ead{orv@tsinghua.edu.cn}
    \author[d]{Sen Qian}
    \author[a,b,c]{Zhe Wang}
    \author[a,b,c]{Shaomin Chen}

    \cortext[cor1]{Corresponding author}
    \affiliation[a]{
        organization={Department of Engineering Physics},
        addressline={Tsinghua University},
        city={Beijing},
        postcode={100084},
        country={China}}
    \affiliation[b]{
        organization={Center for High Energy Physics},
        addressline={Tsinghua University},
        city={Beijing},
        postcode={100084},
        country={China}}
    \affiliation[c]{
        organization={Key Laboratory of Particle \& Radiation Imaging (Tsinghua University)},
        addressline={Ministry of Education},
        country={China}}
    \affiliation[d]{
        organization={Institute of High Energy Physics},
        addressline={Chinese Academy of Sciences},
        city={Beijing},
        postcode={100049},
        country={China}
    }
    \begin{abstract}
        The atomic layer deposition~(ALD) coating lengthens the lifetime of microchannel plates~(MCP),  which are used
        as the electron amplifier of the photomultiplier tubes~(PMT).
        In the Jinping Neutrino Experiment, the newly developed 8-inch MCP-PMT achieves high collection efficiency by
        coating with high secondary emission materials.
        The resulting single electron response~(SER) charge distribution deviates from the Gaussian distribution in large charge regions.
        To understand the nature of the jumbo-charged SER,
        we designed a voltage-division experiment to quantify the dependence of the MCP gain on the energy of incident electrons.
        Combining the relationship with the Furman probabilistic model,
        we reproduced the SER charge spectra by an additional amplification stage on the input electrode of the first MCP.
        Our results favor a Gamma-Tweedie mixture to describe the SER charge spectra of the MCP-PMTs.
    \end{abstract}
\end{frontmatter}
\section{Introduction}\label{sec:Introduction}
The photomultiplier tubes~(PMT) see extensive deployments in particle physics, in particular neutrino experiments.
A PMT comprises a photocathode, an electron multiplier, an anode, and other necessary structural components.~\cite{1955Scintillation}.
Photons from a light source incident on the photocathode follow a Poisson process.
Some of them are converted to photoelectrons~(PE) via the photoelectric effect
and the PEs enter the multiplier~\cite{2016Optimization}.
Those two processes are Bernoulli selections with the probabilities being the quantum efficiency~(QE) and the collection efficiency~(CE).
The PE count~($n_{\mathrm{PE}}$) in a specific time interval follows a Poisson distribution~\cite{1994Absolute}.

The electron amplification is driven by the secondary electron emission~(SEE)
that when an incident particle, electron or ion,
collides with or goes through a solid surface, one or more secondary electrons are emitted~\cite{2016Secondary}.
The average number of the secondaries produced per incident particle is the secondary-emission yield~(SEY, $\delta$).
The energy distribution of the secondary electrons~(\(\mathrm{d}\delta/\mathrm{d}E\)) is related to the energy of the incoming particle,
the incident angle, the target material, etc.~\cite{2002Probabilistic}.
Bruining~\cite{1938Secondary}, Ushio~\cite{1988Secondary} and Jokela~\cite{2012Secondary}
conducted target-shooting experiments using electron guns,
and measured the SEY in the current mode.
L. Olano~\cite{OLANO2020103456} measured the energy distribution \(\mathrm{d}\delta/\mathrm{d}E\) of Kapton, Teflon and Ultem by charging analysis,
and found the energy of the secondaries is much smaller than that of the primary electrons.
Such results are then extrapolated to PMTs~\cite{2012An,2021Effects}.
The low light intensity at which a PMT operates makes the incident electrons discrete.
Therefore, one should be careful when extending the SEY from the current mode to a single electron case, the pulse mode.

\begin{figure}[!htbp]
    \centering
    \includegraphics[width=0.7\textwidth]{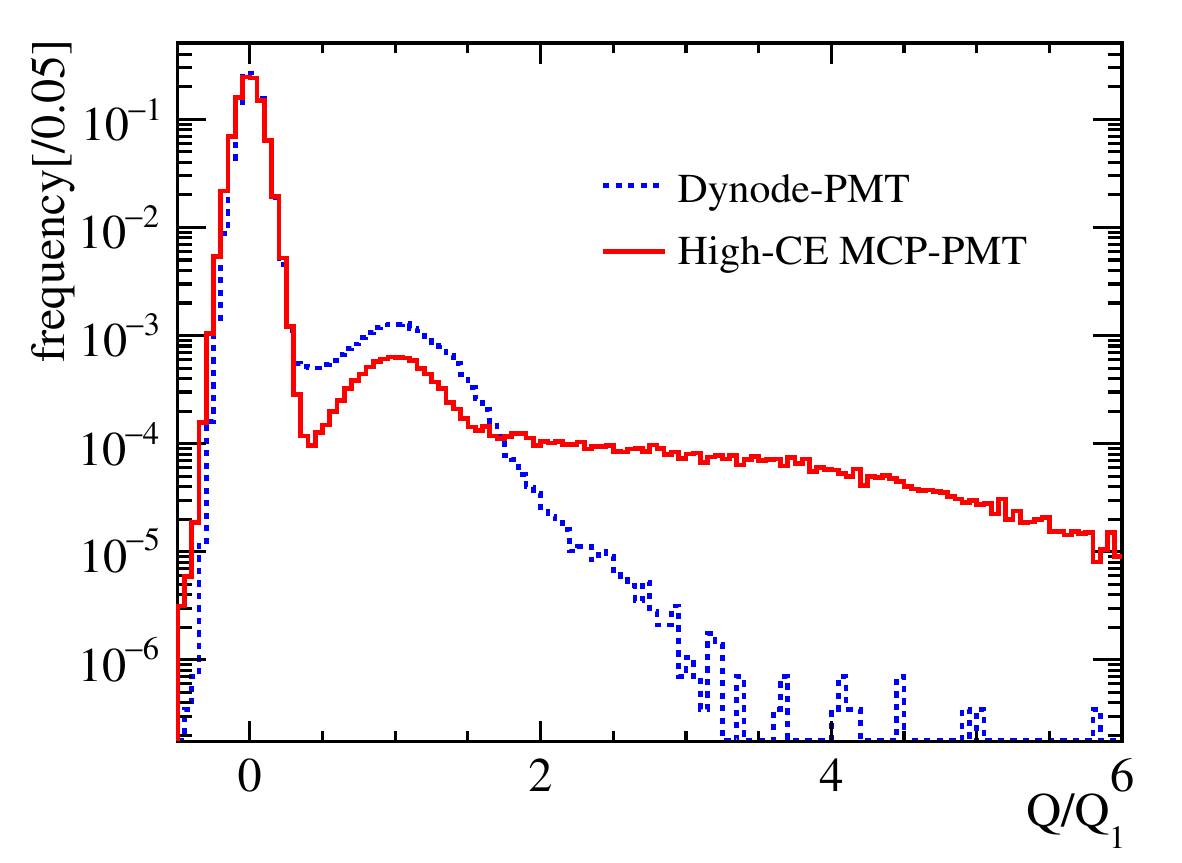}
    \caption{The charge spectrum of the high-CE MCP-PMT GDB-6082~(SN:PM2112-9089F)~(red) and a Dynode-PMT~(blue)~\cite{Zhang:2023ued}.
        The blue histogram consists of the pedestal $Q=0$ and the principal peaks of $Q=Q_1$, while the red histogram includes jumbo charges.}
    \label{fig:spe_sreal}
\end{figure}

After being amplified by the multiplier,
a single PE induces numerous electrons,
which are captured by the anode within a few hundred picoseconds.
The initial energy of the PEs produced at the photocathode is \SI{\sim 1}{eV}~\cite{Nathan1970TheED}.
The incident energy of the PEs arriving at the multiplier is dominated by the potential difference
between the photocathode and the multiplier, therefore the amplifier provides nearly identical gain for the PEs.
Because the total charge of the electrons captured by the anode is typically described by a Gaussian distribution in light of the central limit theorem of probability,
the probability density function~(PDF) of the single electron response~(SER) charge distribution is $f_{\mathcal{N}}(Q; Q_1,\sigma_1^2)$,
where $Q_1$ is the mean charge, and $\sigma_1$ is its standard deviation.
The PE count $n_{\mathrm{PE}}$ follows a Poisson distribution with the probability mass function $P_\pi(n_{\mathrm{PE}};\lambda)$,
where $\lambda$ is the expected PE count at a certain light intensity.
The total charge distribution $f(Q)$ after amplification is a folding of the Poisson distribution and the SER charge distribution~\cite{1994Absolute}.
There are two kinds of background processes:
the low charge one, following the Gaussian distribution~$\mathcal{N}(0,\sigma_0^2)$,
represents a finite-width distribution without PE emitted from photocathode;
and the discrete one with probability $w$, such as the thermoemission and the noise initiated by the incident light, follows the exponential distribution~$\mathrm{Exp(\alpha)}$
with $\alpha$ as the rate of the exponential decrease.
Considering the charge distribution of the two types of background processes being $f_{\mathrm{b}}(Q)$, the overall charge distribution can be expressed in Eq.~(\ref{eq:sreal}):
\begin{equation}
    \begin{aligned}
        f(Q) =  & P_{\pi}(n_{\mathrm{PE}}=0;\lambda)f_{\mathrm{b}}(Q) + P_{\pi}(n_{\mathrm{PE}}>0;\lambda)\bigotimes f_{\mathcal{N}}(Q; n_{\mathrm{PE}}Q_1,n_{\mathrm{PE}}\sigma_1^2) \\
        \approx & \left\{\frac{(1-w)}{\sigma_0 \sqrt{2 \pi}} \exp \left(-\frac{Q^2}{2 \sigma_0^2}\right)
        +w \theta(Q)\times \alpha \exp \left(-\alpha Q\right)\right\} \mathrm{e}^{-\lambda}                                                                                           \\
                & +\sum_{n_{\mathrm{PE}}=1}^{\infty} \frac{\lambda^{n_{\mathrm{PE}}} \mathrm{e}^{-\lambda}}{n_{\mathrm{PE}} !}
        \times \frac{1}{\sigma_1 \sqrt{2 \pi n_{\mathrm{PE}}}}\times
        \exp \left(-\frac{\left(Q-n_{\mathrm{PE}} Q_1\right)^2}{2 n_{\mathrm{PE}} \sigma_1^2}\right)
    \end{aligned}
    \label{eq:sreal}
\end{equation}
where $\theta(Q)$ is the Heaviside function.
When $\lambda$ is less than 0.1,
the probability of observing two or more PEs is less than one-tenth of the probability of observing a single PE.
In this case, the charge distribution will only show the peak of the pedestal ($Q=0$) and the peak of the single PE ($Q=Q_1$) as indicated in the blue histogram in Fig~\ref{fig:spe_sreal}.
After applying some cuts to remove the pedestal, we can obtain an approximate SER charge spectrum
to study the SER charge spectrum divided by $Q_1$ as $Q/Q_1$ to align the gain of different PMTs.

Instead of a large-sized dynode-chain commonly used in PMT,
MCP-PMTs employ MCPs as electron multipliers.
MCP-PMTs are currently in use or planned for neutrino experiments
like the Jiangmen Underground Neutrino Observatory~(JUNO)~\cite{ZHU2020162002} and the Jinping Neutrino Experiment~\cite{Zhang:2023ued},
collider experiments like the Belle II TOP detector~\cite{MATSUOKA2014148} and the PANDA DIRC Cherenkov detector at FAIR~\cite{KRAUSS2023168659}, 
and cosmic ray observatories like the Large High Altitude Air Shower Observatory~\cite{Cao2019UpgradingPT}.
Initially, the fact that the feedback ions cause damage to the photocathode
leads to a lifetime issue of MCP-PMT~\cite{N2006Lifetime}.
A precise thin film deposition technique of atomic layer deposition~(ALD)~\cite{reviewer2, 2012An}
is applied to fabricate MCP-PMTs solving the lifetime issue~\cite{Lehmann:2022ret}.
Lin Chen~et~al.~\cite{2016Optimization} indicated that depositing high SEY materials
such as \ce{Al2O3} via ALD on the input electrode of the first MCP
can enhance the probability of collecting the secondaries to improve CE
to nearly 100\% rather than being constrained by the MCP open area fraction.
This enhancement is later extended to a composite \ce{Al2O3}-\ce{MgO} layer
by Weiwei Cao~et~al.~\cite{cao_secondary_2021} and Zhengjun Zhang~et~al.~\cite{zzj2021Al}
to allow for increased gain, improved single electron resolution,
and a higher peak-to-valley ratio of the MCP-PMTs~\cite{2021Effects}.

In the performance tests to evaluate the 8-inch high-CE MCP-PMT by the Jinping Neutrino Experiment,
\emph{jumbo charges} are found in the SER charge spectra~\cite{Zhang:2023ued},
as shown in the red histogram in Fig~\ref{fig:spe_sreal}.
Similar charges have also been observed in the mass testing of the 20-inch MCP-PMTs at JUNO,
identified as the ``long tail'' in the SER charge distribution~\cite{JUNO:2022hlz}.
Orlov D. A.~et~al.~\cite{reviewer1} reported that for the high-CE MCP-PMTs,
the shape of the pulse height distribution of the single PE events has become wider,
and the probability of a high-energy tail has increased.
H.~Q.~Zhang~et~al.~\cite{2021Gain} used the charge model in Eq.~\eqref{eq:sreal} for the jumbo charges
and recommended an extra gain calibration.
Yuzhen Yang et al.~\cite{2017MCP} conducted a voltage-division experiment to reveal that
the MCP gain for the low-energy electrons is significantly smaller than
that for the high-energy ones.
Thus, the MCP gain for the secondaries is different from that for the PEs entering the channels directly.
The SER charge model in Eq.~\eqref{eq:sreal} is no longer sufficient to accurately calibrate this type of PMT.
The origin of the jumbo charges is necessary for an appropriate SER charge calibration.

In this paper, the Gamma distribution is introduced in Sec.~\ref{gammapossion}.
In Sec.~\ref{sec:see}, a voltage-division experiment is designed to measure the relationship
between MCP gain and the energy of the incident electrons.
Taking into account the SEE model, we elucidate the nature of the jumbo charges
and calculate the total SEY of the \ce{Al2O3}-\ce{MgO} layer when the incident energy is \SI{650}{eV}.
Sec.~\ref{sec:discussion} proposes a Gamma-Tweedie mixture for the MCP-PMT and makes further discussions.
We conclude in Sec.~\ref{sec:conclusion}.

\section{Gamma-Distributed SER charges}\label{gammapossion}
Every multiplication of electrons at the dynodes or MCP channels follows approximately a Poisson distribution~\cite{branchandPoisson}.
A series of such multiplications forms cascaded Poissonians~\cite{1955Scintillation} and is
an example of the branching process~\cite{Bartlett1963TheTO} challenging to perform analytical computations.
Breitenberger~\cite{1955Scintillation} argued that the SER charge spectrum is between the Poisson distribution
and the Gaussian.
Prescott~\cite{polya} proposed the use of a cascaded Polya distribution to characterize the electron multiplication in PMT,
particularly when considering the non-uniformity of the dynode surface.
Kalousis~\cite{2012Calibration,2020A} approximated the Polya distribution as a Gamma one to calibrate PMT
and achieved better results than the Gaussian model in Eq.~\eqref{eq:sreal}.

Instead of the Gaussian containing a small nonphysical tail less than 0,
we choose a Gamma distribution $\varGamma(\alpha, \beta)$
defined by the scale factor $\alpha$ and the rate factor $\beta$, as shown in Eq.~\eqref{eq:gamma}:
\begin{equation}
    \label{eq:gamma}
    \begin{aligned}
        f_\Gamma(x ; \alpha, \beta) = \frac{x^{\alpha-1} e^{-\beta x} \beta^\alpha}{\Gamma(\alpha)} \quad \text { for } x>0 \quad \alpha, \beta>0 \\
    \end{aligned}
\end{equation}
where $\Gamma(\alpha)$ is the Gamma function.
A Gamma distribution is uniquely determined by its expectation \(\alpha/\beta=Q_1\) and variance \(\alpha/\beta^2=\sigma_1^2\)
which can be converted into the Gaussian counterparts in~Eq.~\eqref{eq:sreal}.
The SER charge spectrum based on the Gamma distribution is,
\begin{equation}
    \begin{aligned}
        f(Q) & = & P_{\pi}(n_{\mathrm{PE}}=0;\lambda)f_{\mathrm{b}}(Q) + P_\pi(n_{\mathrm{PE}}>0;\lambda)\bigotimes f_\Gamma(Q;n_{\mathrm{PE}}\alpha, \beta). \\
    \end{aligned}
    \label{eq:Gamma}
\end{equation}

\section{Jumbo Charges through Extra Multiplication}\label{sec:see}

After being discovered, SEE received attention during the widespread application of electronic tubes.
Bruining summarized the methods, findings and applications of SEE in his classic \textit{Physics and Applications of Secondary Electron Emission}~\cite{bruining_physics_1954}.
Baroody~\cite{baroody1950theory} put forward his SEE theory of metals with the assumption that an incident primary
electron interacts only with free electrons in the conduction band,
without considering the variation of secondary emission with the primary energies.
Dekker~et~al.~\cite{dekker1952theory} presented the SEE quantum theory of
the Coulomb interaction between the incident primaries and the lattice electrons.
Wolff~\cite{wolff1954theory} provided the cascade theory for the diffusion, the energy loss and the multiplication of
the secondary electrons within a metal.
Assuming both incident and back-scattered electrons within the target are isotropic, Koichi Kanaya~et~al.~\cite{Kanaya_1978} calculated the SEY from insulators with the ionization potential by setting the valence electron and the back-scattered coefficient
besides the parameter of the free-electron density effect.
Vaughan~\cite{vaughan} formulated the SEY
as a function of impact energy and direction that are used in computer programs, known as the \emph{Vaughan model}.
Furman and Pivi~\cite{2002Probabilistic} developed a mathematically self-consistent Monte Carlo program
to elucidate the SEE phenomenon from solid surfaces usually named the \emph{Furman model}.
This model incorporates the statistical nature of the SEE process
by considering the probability distribution governing the number of the secondaries emitted per incident primary electron.
The energies of secondary electrons are approximated as independent and identically distributed random variables
determined by the material properties and the primary energies.

Early models primarily focused on theoretical explanations of SEE.
The Vauham and Furman models emphasize the Monte Carlo computation instead.
Comparatively, the Furman model strives for physical consistency and better agreement with experiments.
We therefore choose it for more adjustable parameters and higher accuracy.

\subsection{Furman probabilistic model}\label{subsec:fuman}

In the Furman model~\cite{2002Probabilistic}, there are three kinds of secondary electrons.
The first is the back-scattered electron, emitted through elastic scattering on the surface of the target material.
The energy distribution $\mathrm{d}\delta_{\mathrm{bs}}/\mathrm{d}E$ is defined in Eq.~\eqref{eq:backscatter},
where $\delta_{\mathrm{bs}}$ is the yield of the back-scattered electron,
the Heaviside function $\theta(E)$ ensures the $E<E_0$.
$E_0$ is the incident energy of the primary electron,
$\theta_0$ is the incident angle,
and $\sigma_{\mathrm{bs}}$ is an adjustable standard deviation.
\begin{equation}
    \label{eq:backscatter}
    \begin{aligned}
         & \frac{\mathrm{d}\delta_{\mathrm{bs}}}{\mathrm{d}E} =\theta(E) \theta\left(E_0-E\right) \delta_{\mathrm{bs}}\left(E_0, \theta_0\right)
        \frac{2 \exp \left(-\left(E-E_0\right)^2 / 2 \sigma_{\mathrm{bs}}^2\right)}{\sqrt{2 \pi} \sigma_{\mathrm{bs}}
        \operatorname{erf}\left(E_0 / \sqrt{2} \sigma_{\mathrm{bs}}\right)}                                                             \\
    \end{aligned}
\end{equation}

After penetrating the target material, some electrons are inelastically scattered by the atoms and are reflected out to form the second category.
P.~Lenard called the bending of the electron track ``diffusion'',
and the trajectory turning $90^\circ$ as \text {``Rückdiffusion''} in the German literature~\cite{bruining_physics_1954}.
Furman and Pivi adopted this convention to name them as the \emph{rediffused electrons}.
The energy distribution of the rediffused electrons is defined as Eq.~\eqref{eq:rediffused},
where $\delta_{\mathrm{rd}}$ is the yield of rediffused electron,
and $q$ is an adjustable parameter.
\begin{equation}
    \label{eq:rediffused}
    \begin{aligned}
         & \frac{\mathrm{d}\delta_{\mathrm{rd}}}{\mathrm{d}E} =\theta(E) \theta\left(E_0-E\right) \delta_{\mathrm{rd}}\left(E_0, \theta_0\right) \frac{(q+1) E^q}{E_0^{q+1}} \\
    \end{aligned}
\end{equation}

The final and most important kind is the true-secondary electrons.
Upon deeper penetration of electrons into the target material, intricate physical processes ensue,
generating one or more secondaries.
It is the very process of multiplying electrons.
The spectrum is defined as Eq.~\eqref{eq:true}.
\begin{equation}
    \label{eq:true}
    \begin{aligned}
        \frac{\mathrm{d} \delta_{\mathrm{ts}}}{\mathrm{d} E}=  \sum_{n=1}^{\infty}
        \frac{n P_{\mathrm{n, ts}}\left(n; \delta_{\mathrm{ts}}(E_0,\theta_0)\right)
        \left(E / \epsilon_{\mathrm{n}}\right)^{p_{\mathrm{n}}-1} e^{-E / \epsilon_{\mathrm{n}}}}
        {\epsilon_{\mathrm{n}} \Gamma\left(p_{\mathrm{n}}\right) \Upsilon\left(n p_{\mathrm{n}}, E_0 / \epsilon_{\mathrm{n}}\right)}
        \times \Upsilon\left[(n-1) p_{\mathrm{n}},\left(E_0-E\right) / \epsilon_{\mathrm{n}}\right]
    \end{aligned}
\end{equation}
where $\delta_{\mathrm{ts}}(E_0,\theta_0)$
is the yield of the true-secondary electrons when the incident energy is $E_0$ and the incident angle is $\theta_0$,
$\epsilon_{\mathrm{n}}>0$ and $p_{\mathrm{n}}>0$ are the phenomenological parameters.
$\gamma(z,x)$ is the incomplete gamma function,
and $\Upsilon(z,x)=\gamma(z,x)/\Gamma(z)$ is the normalized form satisfying $\Upsilon(0,x)=1$.
$n$, the number of the true-secondary electrons, follows the Poisson distribution~$\mathrm{\pi}(\delta_{\mathrm{ts}}(E_0,\theta_0))$.
$P_{\mathrm{n, ts}}$ is its probability mass function.

\begin{figure}[!htbp]
    \centering
    \includegraphics[width=0.7\textwidth]{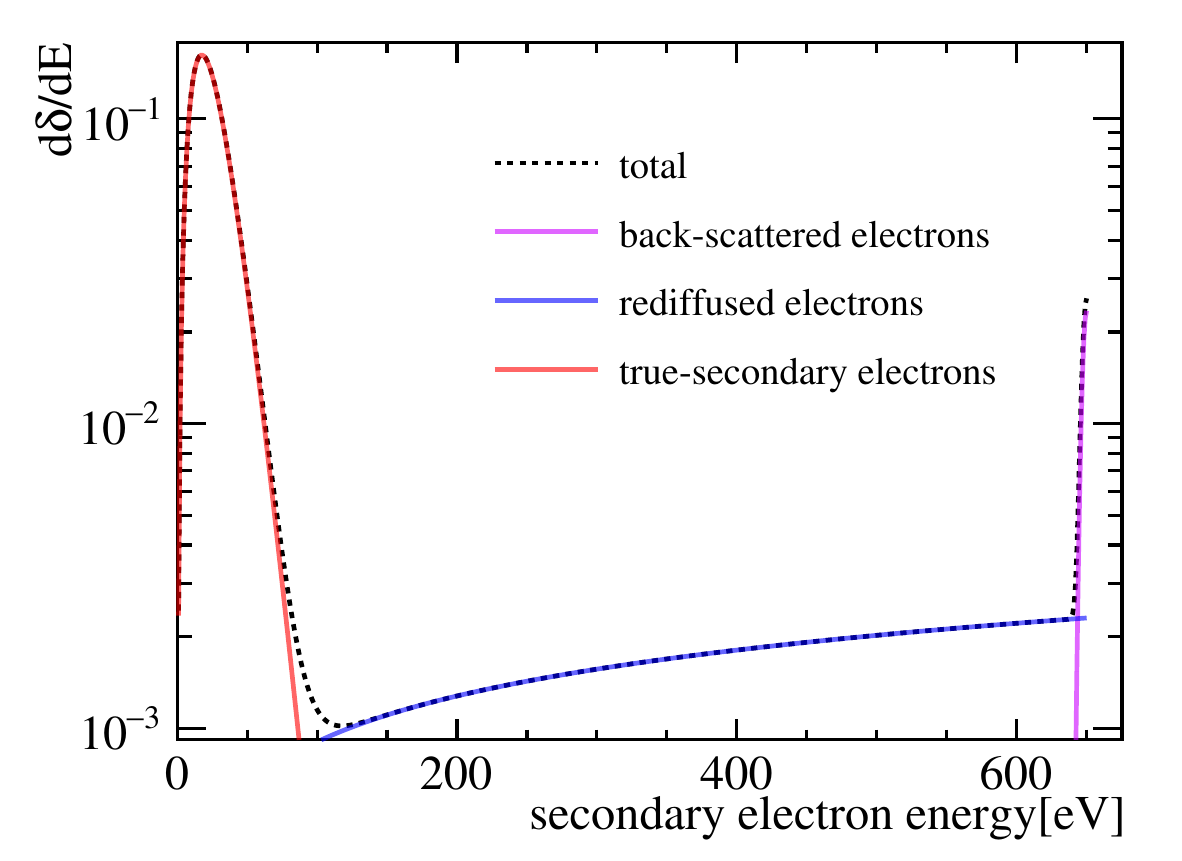}
    \caption{The total energy spectrum of the secondary electrons when the incident energy is \SI{650}{eV}.
        The violet, blue and red lines represent $\mathrm{d}\delta_{\mathrm{bs}}/\mathrm{d}E$, $\mathrm{d}\delta_{\mathrm{rd}}/\mathrm{d}E$ and 
        $\mathrm{d}\delta_{\mathrm{ts}}/\mathrm{d}E$ repectively.
        The black dashed line is $\mathrm{d}\delta/\mathrm{d}E$.}
    \label{fig:SES}
\end{figure}

For illustration in Fig.~\ref{fig:SES}, we set the parameters as
$\delta_{\mathrm{bs}}=0.05$, $\delta_{\mathrm{rd}}=0.5$, $\delta_{\mathrm{ts}}=5$~\cite{2021Effects},
$\theta_0=0^\circ$  and $E_0=$\SI{650}{eV},
and the total spectrum is
$\mathrm{d}\delta/\mathrm{d}E=\mathrm{d}\delta_{\mathrm{bs}}/\mathrm{d}E+\mathrm{d}\delta_{\mathrm{rd}}/\mathrm{d}E+\mathrm{d}\delta_{\mathrm{ts}}/\mathrm{d}E$.
The energies of the secondaries are usually less than \SI{100}{eV} when the incident energy~$E_0$ is \SI{650}{eV}.

\subsection{An extra multiplication mode}
The MCP-PMT under study uses a chevron stack of two MCPs as the electron multiplier.
As shown in Fig.~\ref{fig:circuit}, an \ce{Al2O3}-\ce{MgO}-\ce{Al2O3} layer~\cite{zzj2021Al} is deposited
on the channel surface of the lead glass body
as well as on the entrance electrode M1 of the first MCP through the ALD technology.
There are two alternative routes of amplification for every PE:
the \textit{channel mode} where the PE directly enters a channel,
and the \textit{surface mode} where the secondaries from $\mathrm{M}1$ enter the MCP channels under the focusing electric field.
The selection of these two routes is a Bernoulli trial~\cite{1955Scintillation}.

The MCP gain for those low-energy secondaries in the surface mode is substantially
smaller than that for the primary PEs in the channel mode~\cite{2012An}.

\subsection{Voltage-division Experiment}\label{sec:gain}
The dependence of the MCP gain for an electron on its incident energy at the
channel entrance is crucial to the origin of jumbo charges.
We designed a voltage-division experiment to measure such a relationship.

\begin{figure}[!ht]
    \centering
    \includegraphics[width=\linewidth]{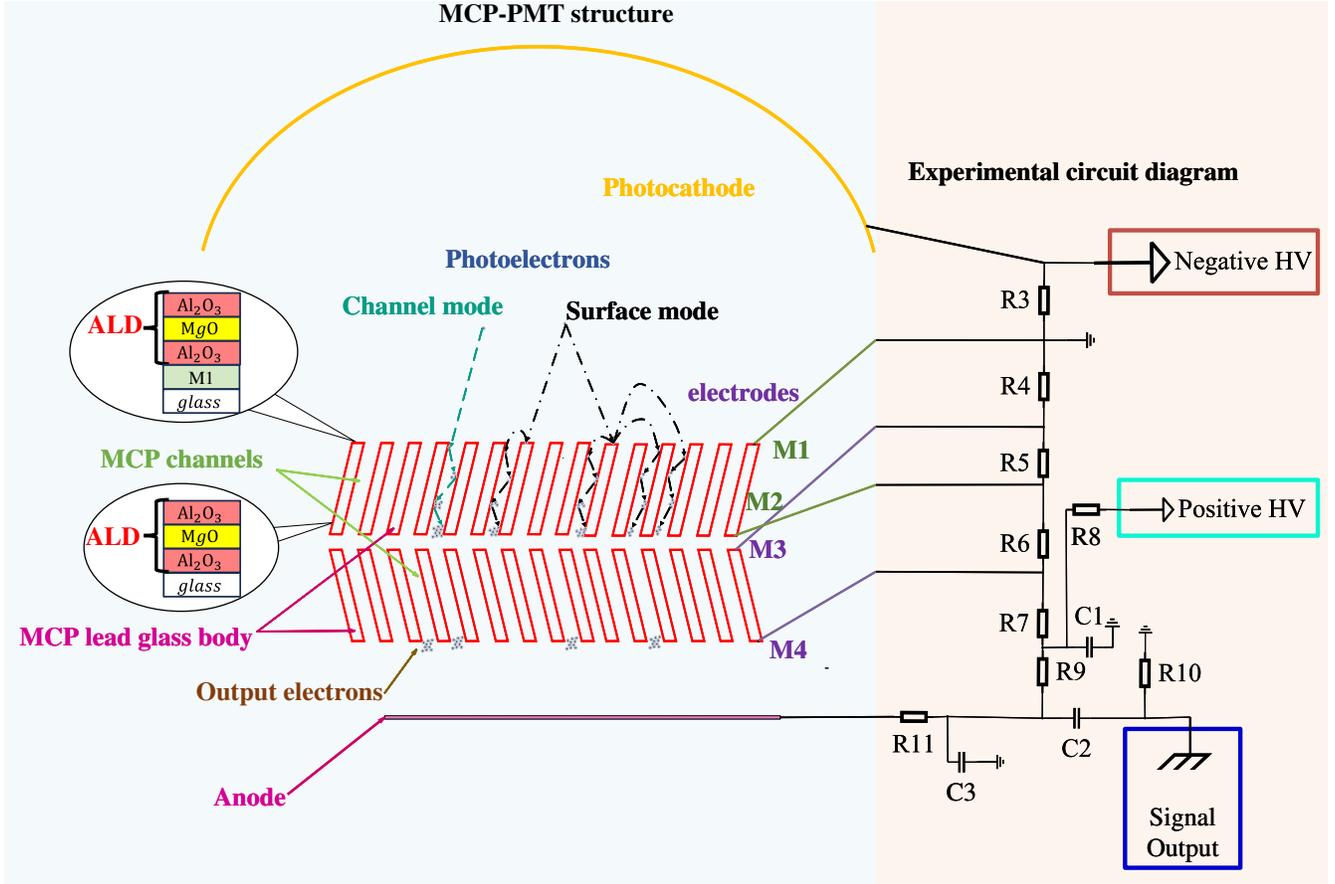}
    \caption{
        MCP-PMT structure: $\mathrm{M}1$ and $\mathrm{M}3$ are the input electrodes of MCPs, $\mathrm{M}2$ and $\mathrm{M}4$ are the output electrodes,
        and the four electrodes provide the potential differences during operation.
        The PEs directly enter the channels~(channel mode) or hit $\mathrm{M}1$ to produce secondary electrons
        that enter the channels later~(surface mode)~\cite{2016Optimization}. After entering the MCP channel, the electron collides
        with the channel wall many times and is amplified in a series of such multiplications~\cite{1955Scintillation}.
        The experimental circuit diagram:
        supply the photocathode with the negative HV and the two MCPs with the positive HV.
        Use an RC filtering and shaping circuit
        to convert the collected electrons at the anode into waveform output.
        The gap voltage between MCPs is added between $\mathrm{M}2$ and $\mathrm{M}3$.}
    \label{fig:circuit}
\end{figure}

As shown in Fig~\ref{fig:circuit}, we utilized a positive high-voltage power supply~(positive HV) to stabilize the potentials applied
to the MCPs through the circuit~\cite{Luo:2023jdf}.  In parallel, we took a negative high-voltage power supply~(negative HV) for varying the electric
potential difference between the photocathode and $\mathrm{M}1$ to get PEs at different incident energies.
Compared to the experiment of Yuzhen Yang~et~al.~\cite{2017MCP} where the potentials of all the electrodes M1-4
are controllable, our design is a simplified adaptation to only tune the energies of the PEs with commercially available HV products.

We used a picosecond laser with a wavelength of \SI{405}{nm} to illuminate the MCP-PMT at a frequency of \SI{1}{kHz},
and used the laser signal as a trigger to capture waveform data.
To obtain the single PE,  we adjusted the intensity of the laser until the occupancy was below 0.1.
We deployed a 10-bit oscilloscope~(HDO9000 with HD1024 Technology)~\cite{teledynelecroy} to capture the \SI{100}{ns} waveform
with a sampling rate of \SI{40}{GS/s} and a range of [-20, 60]~\si{mV}.

\begin{figure}[!ht]
    \centering
    \begin{subfigure}[b]{0.48\textwidth}
        \centering
        \includegraphics[width=\textwidth]{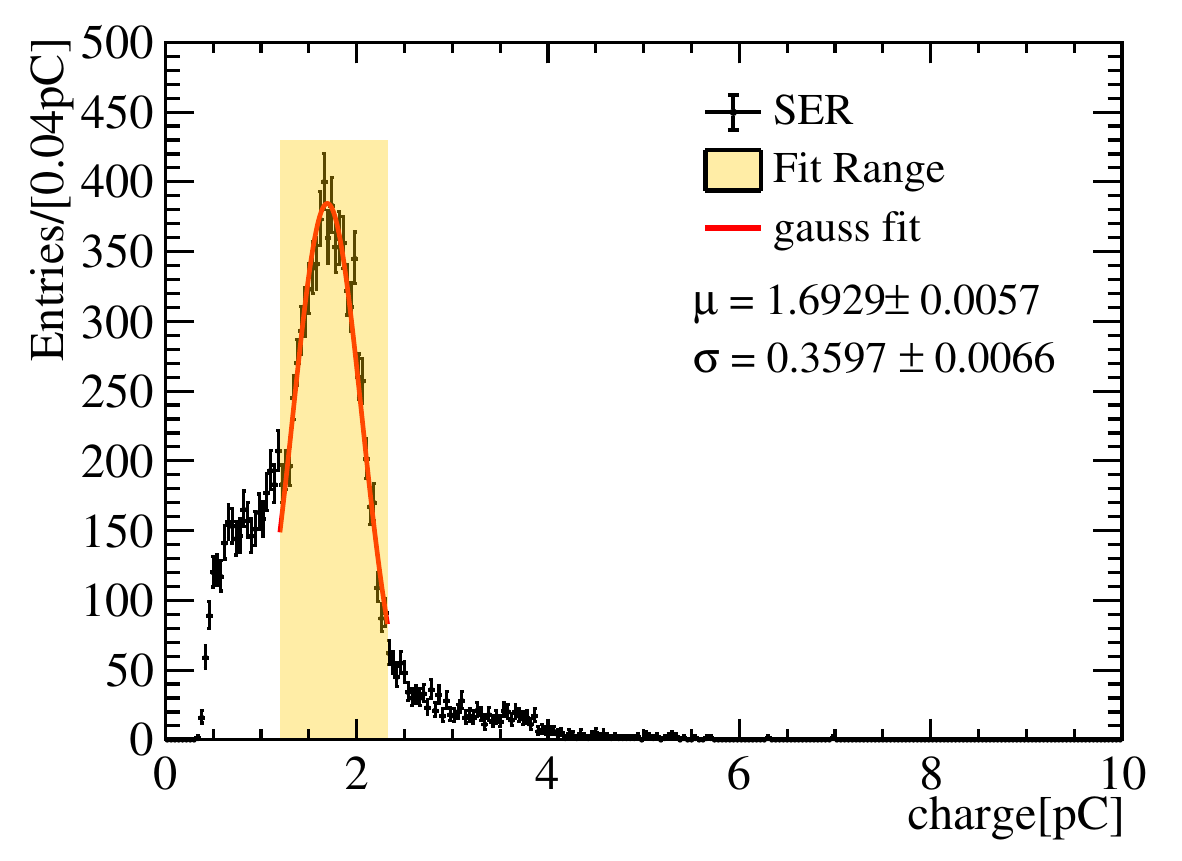}
        \caption{}
        \label{fig:gain_noald}
    \end{subfigure}
    \hfill
    \begin{subfigure}[b]{0.48\textwidth}
        \centering
        \includegraphics[width=\textwidth]{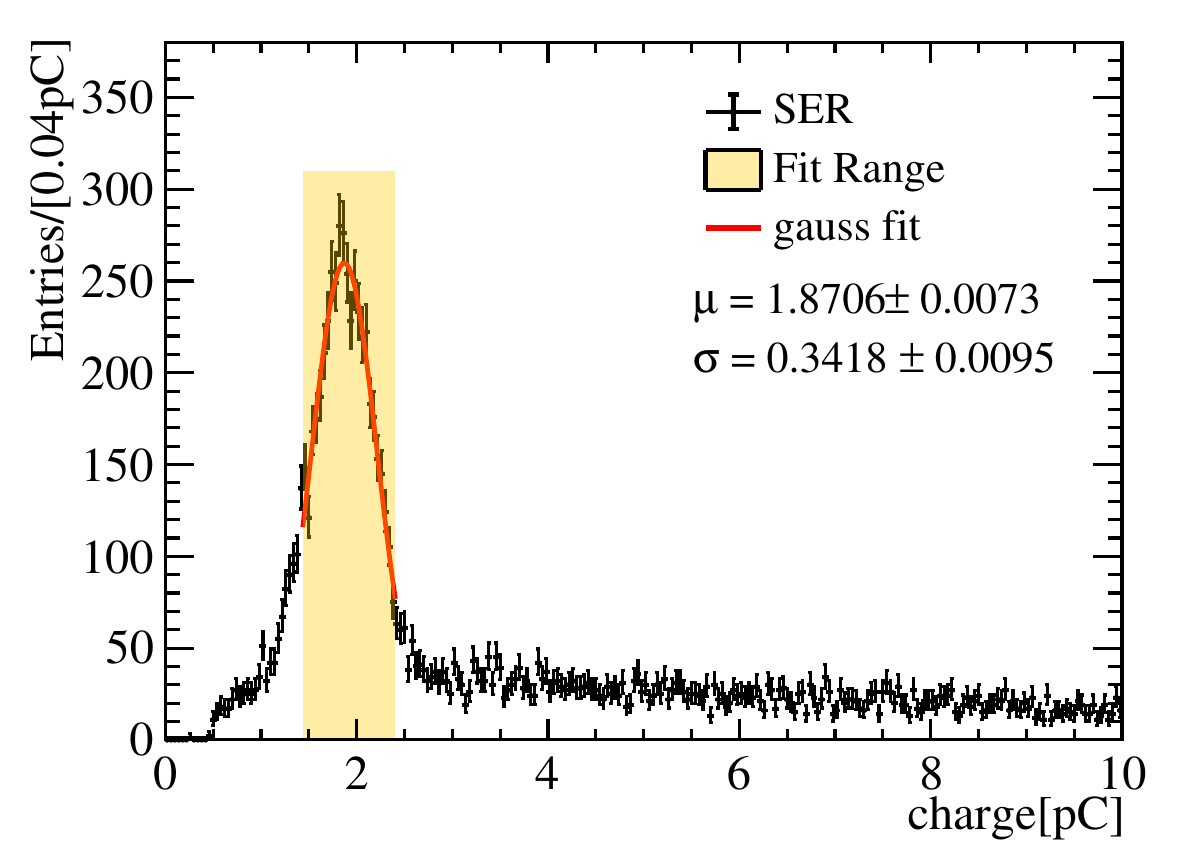}
        \caption{}
        \label{fig:gain_ald}
    \end{subfigure}

    \caption{Fit of the charge spectrum of the MCP-PMT without (\subref{fig:gain_noald}) and with (\subref{fig:gain_ald})
        \ce{Al2O3}-\ce{MgO} deposite on $\mathrm{M}1$.
        We observed that (\subref{fig:gain_noald}) does not exhibit jumbo charges. 
        The yellow areas are incident energy-dependent intervals and
        the red lines are fitting results of the Gaussian functions in the intervals. 
    }
    \label{fig:gain_fit}
\end{figure}

We obtained the gain of the MCP-PMTs at different energies of the incident electrons by fitting the Gaussian on the charge distribution,
and conducted the same experiment on two MCP-PMTs with~(Fig.~\ref{fig:gain_ald}) and without~(Fig.~\ref{fig:gain_noald})
\ce{Al2O3}-\ce{MgO} deposited on $\mathrm{M}1$ to contrast the effect of the surface mode.
The positive voltages for the MCP-PMT with and without \ce{Al2O3}-\ce{MgO} on M1 are +\SI{1205}{V} and  +\SI{1240}{V}, respectively.
Since the initial energies of the PEs are \SI{\sim 1}{eV}~\cite{Nathan1970TheED}
and the systematic error of the negative HV itself is within \SI{2}{V},
the incident energies~($E_0$) are defined as the energies acquired by the PEs in the electric field,
numerically equal to the potential difference between the photocathode and M1, with an error of $\pm$\SI{2}{eV}.
We scanned the MCP gain and measured it every \SI{10}{eV} when $10\leqslant E_0<100$~\si{eV}, every \SI{20}{eV} when $100\leqslant E_0<200$~\si{eV}
and every \SI{50}{eV} when $200<E_0\leqslant 650$~\si{eV}.
For the MCP-PMT with \ce{Al2O3}-\ce{MgO} deposite on M1, our scan range is $10\leqslant E_0\leqslant 600$~\si{eV},
and for the MCP-PMT without, the range is $10\leqslant E_0\leqslant 680$~\si{eV}.

The charges of the captured waveforms were measured with \emph{fast stochastic matching pursuit}~(FSMP)~\cite{Xu_2022,Wang_2024},
which suppresses the interference of electronic noise to give accurate charge spectra under a wide range of gain.
Due to FSMP's ability to count PEs, the charge would be 0 when $n_{\mathrm{PE}}=0$
and the pedestal is cleanly cut out in the output charge distribution.
In Fig.~\ref{fig:gain_fit}, the main peaks are attributed to the channel mode.
The jumbo charges from the surface mode are to the right and deficient amplifications
are to the left of the main peaks.
Since the small contribution of the secondaries from the surface mode for the MCP-PMT without ALD coating on $\mathrm{M}1$, 
there is no jumbo charge in the charge spectrum as shown in Fig.~\ref{fig:gain_noald}. 
To obtain the gain of electrons directly entering the channels,        
only the main peak is fitted to exclude the influence of the surface mode.

Before a detailed fit, we performed a rough fit to obtain approximate values of $\mu_0$ and $\sigma_0$ of the charge distribution.
Subsequently, based on the obtained $\mu_0$ and $\sigma_0$, we provided initial values and ranges for the fit.
The fit ranges were determined from the incident energies of the PEs,
when $E_0>100$~\si{eV}, it was $[\mu_0-1.3\sigma_0, \mu_0+1.6\sigma_0]$;
when $30<E_0\leqslant 100$~\si{eV}, $[\mu_0-0.8\sigma_0, \mu_0+1.6\sigma_0]$;
and when $E_0\leqslant 30$~\si{eV}, $[\mu_0-1.5\sigma_0, \mu_0+1.8\sigma_0]$.
It is sufficient to extract the mean charge $\mu(E_0)$ and the standard deviation $\sigma(E_0)$ 
of the channel-mode main peak to measure the MCP gain for electrons at different energies.

\begin{figure}[!ht]
    \centering
    \begin{subfigure}[b]{0.325\textwidth}
        \centering
        \includegraphics[width=\textwidth]{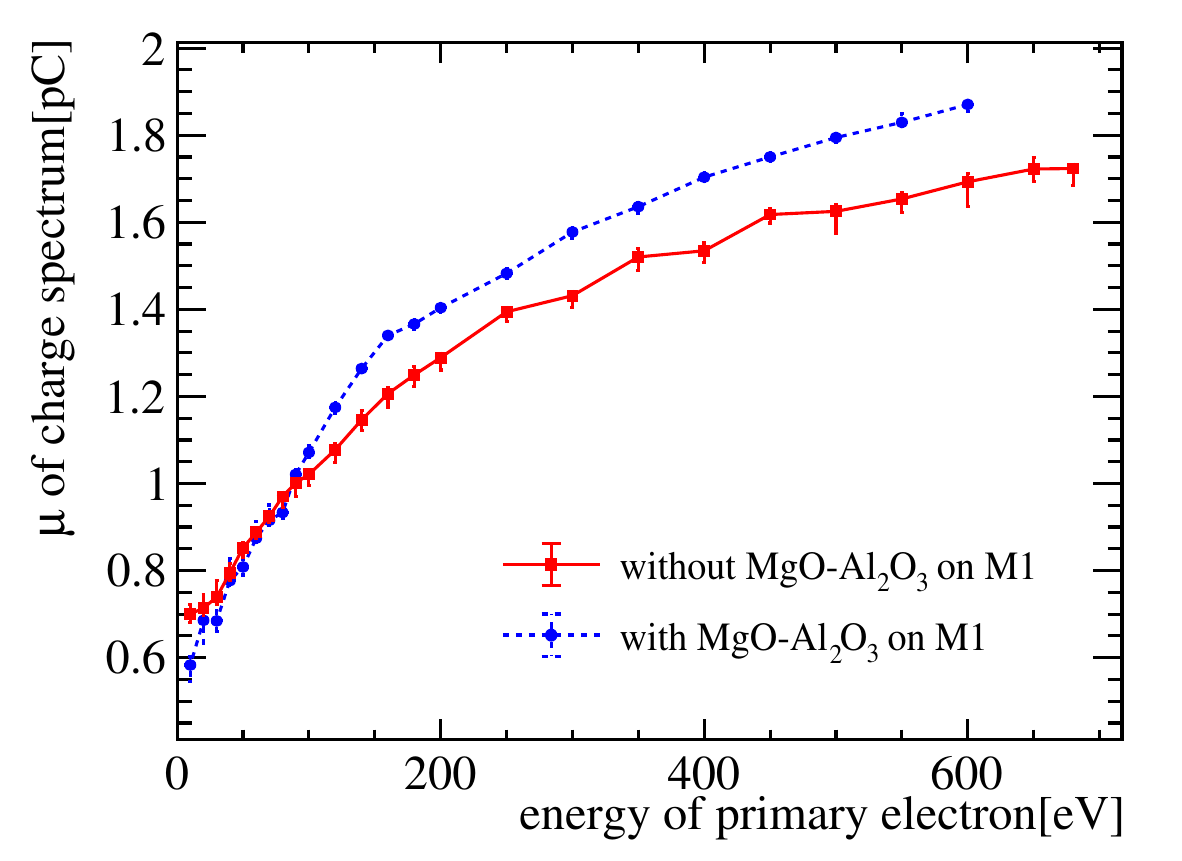}
        \caption{}
        \label{fig:gain}
    \end{subfigure}
    \hfill
    \begin{subfigure}[b]{0.325\textwidth}
        \centering
        \includegraphics[width=\textwidth]{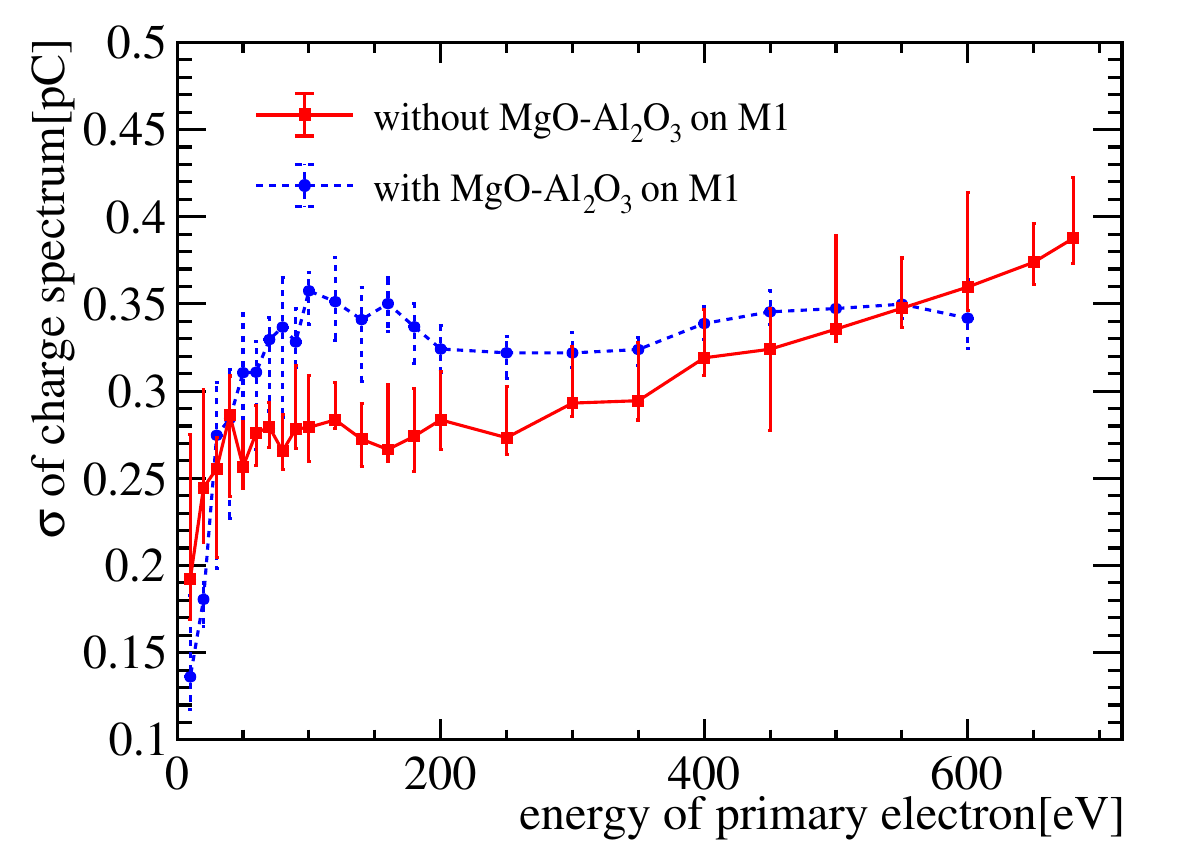}
        \caption{}
        \label{fig:sigma}
    \end{subfigure}
    \hfill
    \begin{subfigure}[b]{0.325\textwidth}
        \centering
        \includegraphics[width=\textwidth]{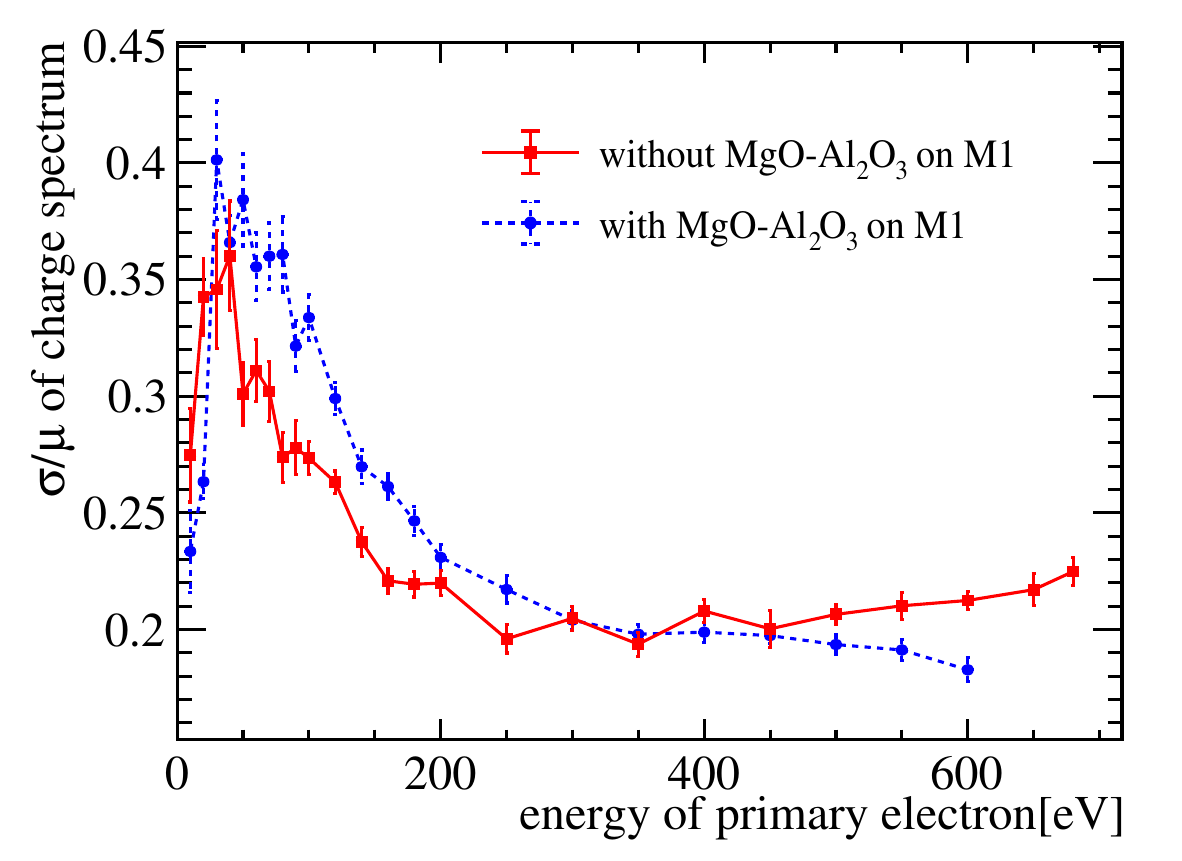}
        \caption{}
        \label{fig:sigmamu}
    \end{subfigure}
    \caption{(\subref{fig:gain}) The mean \(\mu\) increases as the incoming electron energy~$E_0$ increases.
        (\subref{fig:sigma}) The standard deviation \(\sigma\) changes with energy.
        The MCP-PMT with \ce{Al2O3}-\ce{MgO} deposite on $\mathrm{M}1$~(the red line) shows a similar variation trend to the one without~(the blue line).
        (\subref{fig:sigmamu}) The resolution $\sigma/\mu$ increases between 0-\SI{50}{eV}, decreases between 50-\SI{400}{eV},
        and after \SI{400}{eV}, there is a slight decrease for those with \ce{Al2O3}-\ce{MgO} deposite on $\mathrm{M}1$ ~(the blue line) and a slight increase for those without~(the red line).
    }
    \label{fig:gaintest}
\end{figure}

After fitting with Gaussians, we interpolate and extrapolate linearly to obtain the relations of $\mu(E_0)$ and $\sigma(E_0)$ in Fig.~\ref{fig:gaintest}.
The difference in the relations of MCP-PMTs with and without \ce{Al2O3}-\ce{MgO} deposite on $\mathrm{M}1$ comes from the influence of the charge contributed from the suface mode.
When \(E_0 < \SI{200}{eV}\), \(\mu(E_0)\) rapidly increases.
As \(E_0 > \SI{200}{eV}\), \(\mu(E_0)\) gradually stabilizes.
The $\sigma(E_0)$ is overall increasing similar to $\mu(E_0)$ but sees a drop around \SI{200}{eV}.
Similar trend of \(\mu(E_0)\) is reported by Yuzhen Yang~et~al.~\cite{2017MCP}.
In our case the best relative resolution \(\sigma/\mu\) is at around \SI{600}{eV} and
Yuzhen Yang's results suggested \SI{200}{eV}.
Weiwei Cao~et~al.\cite{cao_secondary_2021} found that the SEY of \ce{Al2O3}-\ce{MgO} 
increases with the incident energy in 100-\SI{600}{eV}.
Even though the film structure and thickness we used are different,
we can still roughly determine that
the trend of $\sigma/\mu$ we obtained is reasonable 
based on the variation curves of the SEY of \ce{Al2O3} and \ce{MgO} with energy.

\subsection{Charge-Spectra Decomposition}\label{sec:convolution}
\begin{figure}[!ht]
    \centering
    \includegraphics[width=\linewidth]{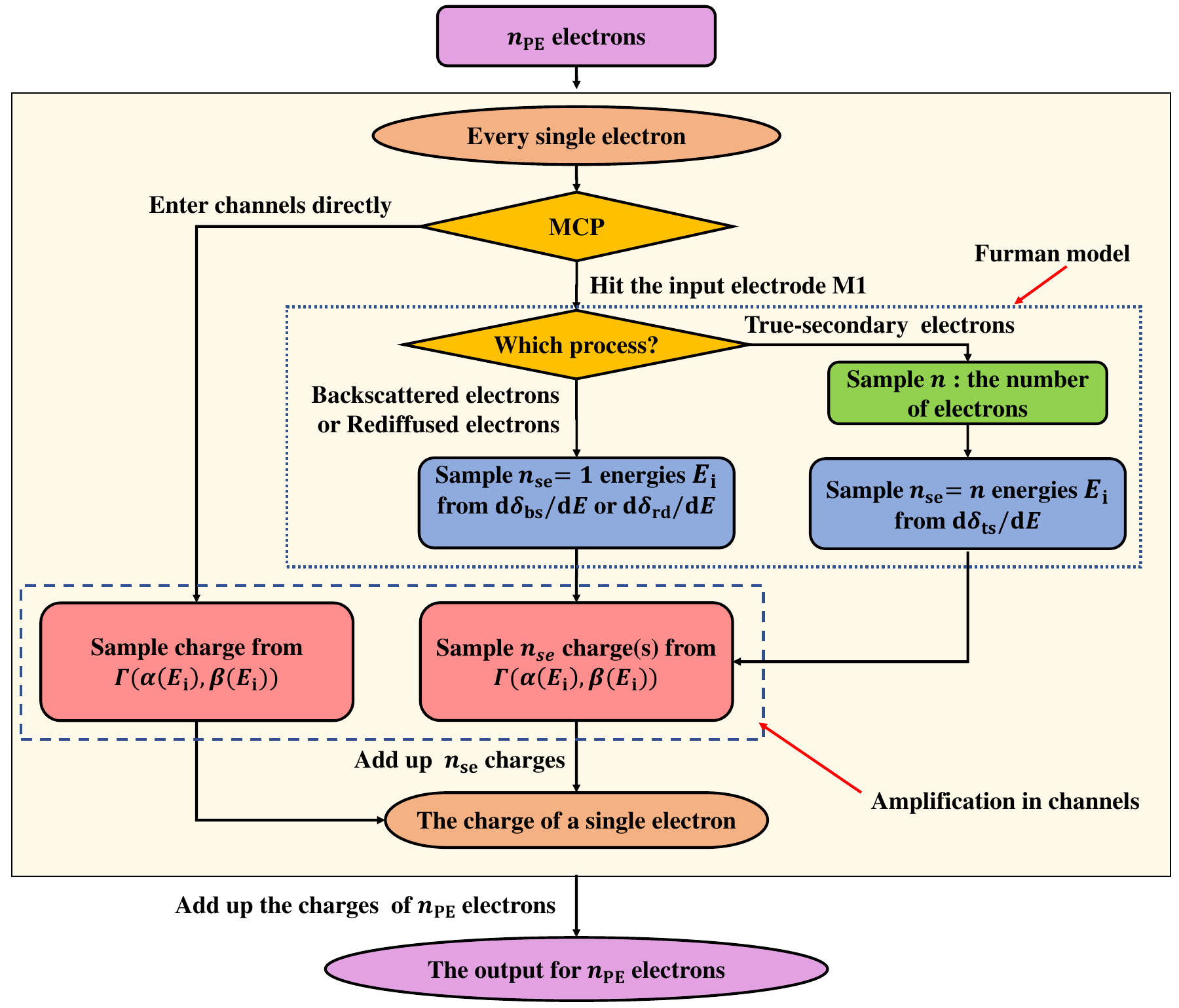}
    \caption{The flowchart of Monte Carlo to compute the charge spectrum. The output charge consists of $n_\mathrm{PE}$ SER charges. The PEs in the channel mode enter channels directly, while PEs in the surface mode hit on the input electrode. The energies of the $n_\mathrm{se}$ secondaries in the surface mode are sampled accroding to the Furman model. The amplification in channels is modeled by the incident energy-dependent Gamma distribution.
    }
    \label{fig:process}
\end{figure}

The Furman model in Sec.~\ref{subsec:fuman} predicts the energies of the secondaries
and our voltage-division experiment in Sec.~\ref{sec:gain} measured the relationship between the 
MCP gain and the incident energies of the electrons.
We follow the flowchart in Fig.~\ref{fig:process} to calculate the charge distribution
by Monte Carlo~(MC)~\cite{1951Various}.
In the study, the laser was directed at the top of the MCP-PMT,
and the PEs originated from the top, resulting in an incident angle~$\theta_0=0^\circ$ when they hit M1.
The complex amplification process in the channels is described by
the incident energy-dependent Gamma distributions $\varGamma(\alpha(E), \beta(E))$ described in Sec.~\ref{gammapossion}. The $\alpha(E)$ and $\beta(E)$ are estimated with the relations of $\mu(E)$ and $\sigma(E)$ of MCP-PMT without ALD coating on the input electrode, which eliminates the influence of the surface mode.
Taking into account of the light intensity, we repeatedly sample $n_{\mathrm{PE}}$ from the Poisson distribution
and sum up $n_{\mathrm{PE}}$ SER charges for the output to get a spectrum.

For sampling an SER charge, we assign the probabilities of the channel and surface
modes as $p$ and $1-p$ to do a Bernoulli trial. The SER charge spectrum $f_{\text{MCP-PMT}}(Q)$ is
\begin{equation}
    \label{eq:convolution}
    f_{\text{MCP-PMT}}(Q) = p f_{\mathrm{ch}}(Q)+(1-p) f_{\mathrm{surf}}(Q)
\end{equation}
where $f_{\mathrm{ch}}(Q)$ and $f_{\mathrm{surf}}(Q)$ are the charge distributions of the channel and surface modes.
$f_{\mathrm{ch}}(Q)$ is set to \(f_\Gamma(Q; \alpha(E_0), \beta(E_0))\), with the incident energy being \SI{650}{eV}.
The factors \(\alpha(E_0), \beta(E_0)\) are converted from $\mu(E_0)$ and $\sigma(E_0)$ without ALD coating in Fig.~\ref{fig:gaintest}.

The \(f_{\mathrm{surf}}(Q)\) is divided into three components by the Furman model
corresponding to Eqs.~(\ref{eq:backscatter})--(\ref{eq:true}),
$f_{\mathrm{bs}}(Q)$ for the back-scattered electrons,
$f_{\mathrm{rd}}(Q)$ for the rediffused electrons,
and $f_{\mathrm{ts}}(Q)$ for the true-secondary electrons.
\begin{equation}
    \label{eq:surface_3}
    \begin{aligned}
        f_{\mathrm{surf}}(Q) & = p_{\mathrm{bs}} f_{\mathrm{bs}}(Q) + p_{\mathrm{rd}} f_{\mathrm{rd}}(Q) + (1- p_{\mathrm{bs}} - p_{\mathrm{rd}}) f_{\mathrm{ts}}(Q)                     \\
                             & = \delta_{\mathrm{bs}} f_{\mathrm{bs}}(Q) + \delta_{\mathrm{rd}} f_{\mathrm{rd}}(Q) + (1- \delta_{\mathrm{bs}} - \delta_{\mathrm{rd}}) f_{\mathrm{ts}}(Q) \\
    \end{aligned}
\end{equation}
where $p_{\mathrm{bs}}$ and $p_{\mathrm{rd}}$ are the mixture ratios
determined by the composition and thickness of surface emissive material
that varies among the PMTs.  Furman and Pivi~\cite{2002Probabilistic} assume that
only one electron is emitted in back-scattered mode and rediffused mode
so that \(\delta_{\mathrm{bs}} = p_{\mathrm{bs}}\) and \(\delta_{\mathrm{rd}} = p_{\mathrm{rd}}\).
In the calculation, we specify $\delta_{\mathrm{rd}}=0.09$ and $\delta_{\mathrm{bs}}=0.01$.
The energy of the back-scattered electron is nearly equal to that of the PEs in the channel mode,
and so is the MCP gain for them.
The energy of the rediffused electron
is lower -- from \SIrange{100}{600}{eV}, causing the charge after MCP multiplication
to be slightly smaller thanks to the relatively slow increase of gain in that range in Fig.~\ref{fig:gain}.
Either contributes a single electron and is practically indistinguishable from
the channel mode in the charge spectra. Such a degeneracy is summarized in Eq.~\eqref{eq:gammaTweedie}:
\begin{equation}
    \label{eq:gammaTweedie}
    \begin{aligned}
        f_{\text{MCP-PMT}}(Q) & = p f_{\mathrm{ch}}(Q) + (1-p) f_{\mathrm{surf}}(Q)                                                                                                                                    \\
                              & = p f_{\mathrm{ch}}(Q)+(1-p) [\delta_{\mathrm{bs}} f_{\mathrm{bs}}(Q) + \delta_{\mathrm{rd}} f_{\mathrm{rd}}(Q) + (1- \delta_{\mathrm{bs}} - \delta_{\mathrm{rd}}) f_{\mathrm{ts}}(Q)] \\
                              & = [p + (1-p)(\delta_{\mathrm{bs}} + \delta_{\mathrm{rd}})] f_{\mathrm{ch}}(Q) + (1-p)(1-\delta_{\mathrm{bs}} - \delta_{\mathrm{rd}})f_{\mathrm{ts}}(Q)
    \end{aligned}
\end{equation}
where the spectra $f_{\mathrm{ch}}(Q)$, $f_{\mathrm{rd}}(Q)$ and $f_{\mathrm{bs}}(Q)$ are
merged into $f_{\mathrm{ch}}(Q)$.

Nevertheless, Eq.~(\ref{eq:gammaTweedie}) is incomplete.
We should consider the case when the secondaries hit the MCP surface again.
The round trip does not inject extra energy.
A back-scattered or rediffused secondary gets amplified essentially in the same way
as a primary PE, while a true-secondary electron has too low an energy to multiply again.
Therefore, \(p_0\), the net contribution to \(f_{\mathrm{ch}}(Q)\), is a geometric series
\begin{equation}
    \label{eq:p0}
    p_0 \coloneqq p \sum_{{\mathrm{i}}=0}^\infty [(1-p) (\delta_{\mathrm{bs}} + \delta_{\mathrm{rd}})]^{\mathrm{i}} = \frac{p}{1 - (1-p) (\delta_{\mathrm{bs}} + \delta_{\mathrm{rd}})}
\end{equation}
and \(f_{\mathrm{ts}}(Q)\) gets \(\frac{(1-p)(1-\delta_{\mathrm{bs}} - \delta_{\mathrm{rd}})}{1 - (1-p) (\delta_{\mathrm{bs}} + \delta_{\mathrm{rd}})}\) or \(1-p_0\).
Eq.~(\ref{eq:gammaTweedie}) is remarkably reduced to
\begin{equation}
    \label{eq:1}
    f_{\text{MCP-PMT}}(Q) = p_0 f_{\mathrm{ch}}(Q) + (1-p_0) f_{\mathrm{ts}}(Q).
\end{equation}

In the case of the true-secondary electrons, their count $n$ follows a Poissonian. The sum of the sampled $n$ charges
serves as the output \(Q_{\mathrm{ts}}\),
\begin{equation}
    \label{eq:ts_all}
    \begin{aligned}
         & Q_{\mathrm{ts}} = \sum_{\mathrm{i}=1}^{n} Q_{\mathrm{i}}        \\
         & n \sim \mathrm{\pi}(\delta_{\mathrm{ts}}')  \\
         & Q_{\mathrm{i}} \sim \varGamma[\alpha(E_{\mathrm{i}}), \beta(E_{\mathrm{i}})] \\
    \end{aligned}
\end{equation}
where \(E_{\mathrm{i}}\) are sampled from Eq.~(\ref{eq:true}). 
$\alpha(E_{\mathrm{i}})$ and $\beta(E_{\mathrm{i}})$ are converted from 
\(\mu(E_{\mathrm{i}})\) and \(\sigma(E_{\mathrm{i}})\) of Fig.~\ref{fig:gaintest}. 
For completeness, \(\delta_\mathrm{ts} \coloneqq (1-\delta_{\mathrm{bs}} - \delta_{\mathrm{rd}})\delta_{\mathrm{ts}}'\) 
is the electric-current ratio of the true-secondary electrons to that of the primary.

\begin{figure}[!htbp]
    \begin{subfigure}{0.47\linewidth}
        \includegraphics[width=\textwidth]{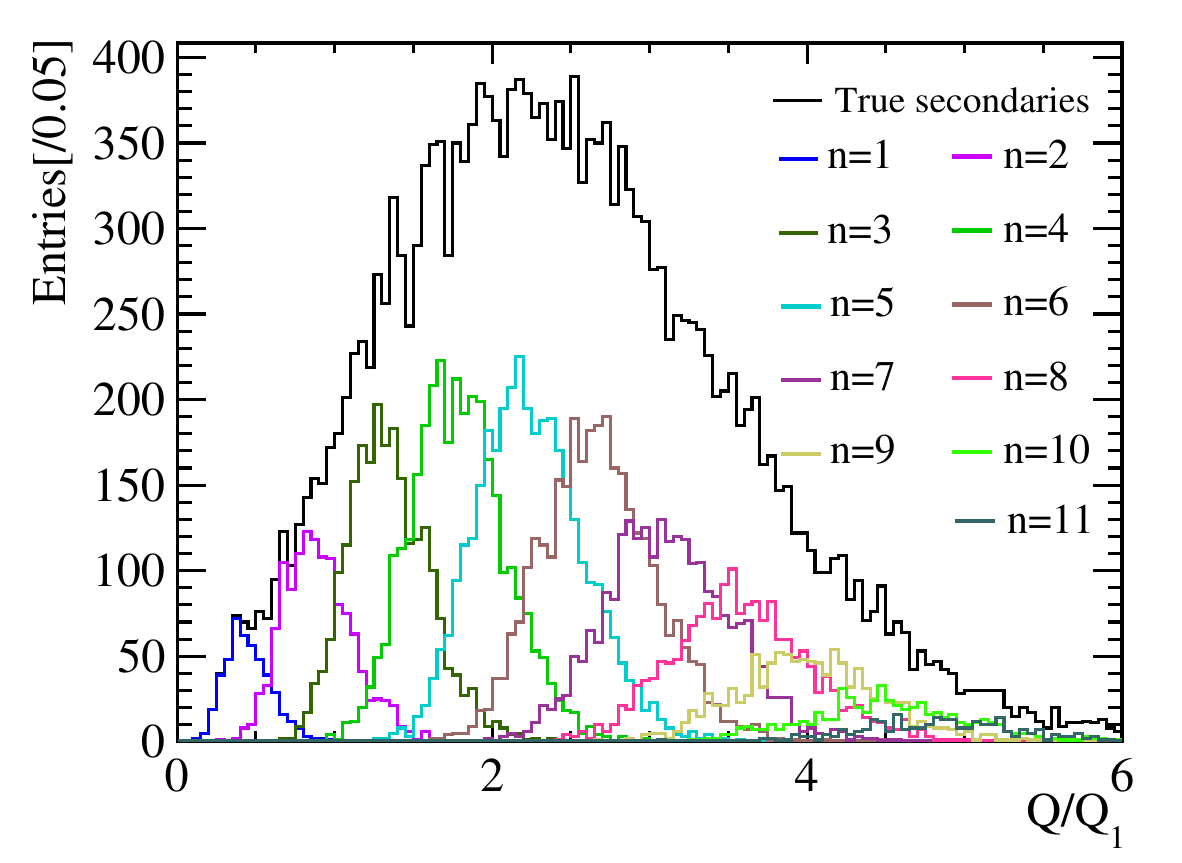}
        \caption{}
        \label{fig:true_n}
    \end{subfigure}
    \hfill
    \begin{subfigure}{0.47\linewidth}
        \includegraphics[width=\textwidth]{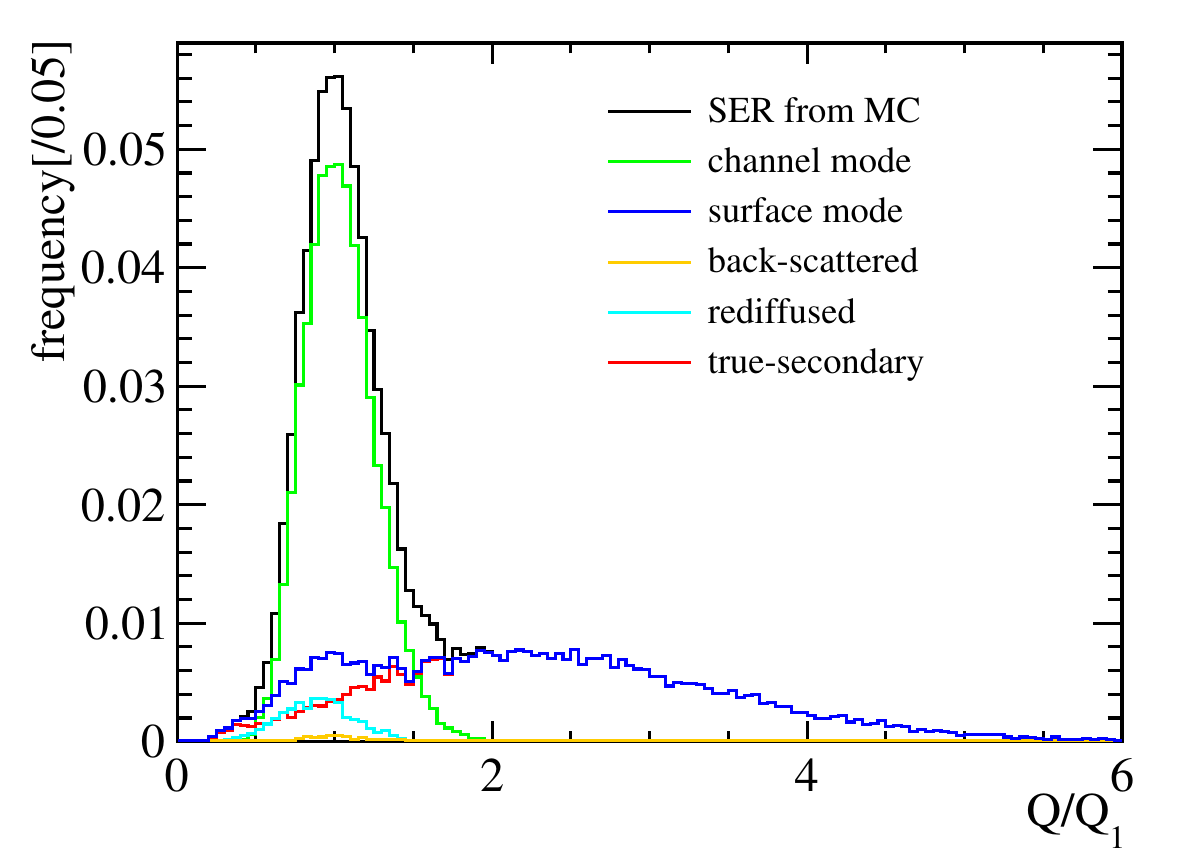}
        \caption{}
        \label{fig:allmode}
    \end{subfigure}
    \caption{(\subref{fig:true_n}) The charge distribution of the true-secondary electrons mode
        in the MC calculation when $\delta_{\mathrm{ts}}'=5.5$ and $p_0=0.55$.
        The black histogram gives the sum of all the distributions.
        (\subref{fig:allmode}) The charge distribution formed in the channel mode is concentrated around the main peak,
        while the tail portion is mainly generated by the true-secondary electrons in the surface mode.}
\end{figure}

The charge spectrum of different $n$ is shown in Fig.~\ref{fig:true_n}.
Due to the lower energies of the secondaries, their charges are smaller.
It is challenging to distinguish each charge formed at the anode,
as multiple secondary electrons enter the MCP channels simultaneously.
Bigger $n$ results in a larger charge.

A typical decomposition of the SER charge spectra is shown in Fig.~\ref{fig:allmode}.
The jumbo charges, also known as the ``long tail'' are contributed by the true secondaries
from the surface mode.

\subsection{Parameter Extraction from Data}\label{subsec:chitest}
\begin{figure}[!htbp]
    \centering
    \begin{subfigure}{0.47\textwidth}
        \centering
        \includegraphics[width=\linewidth]{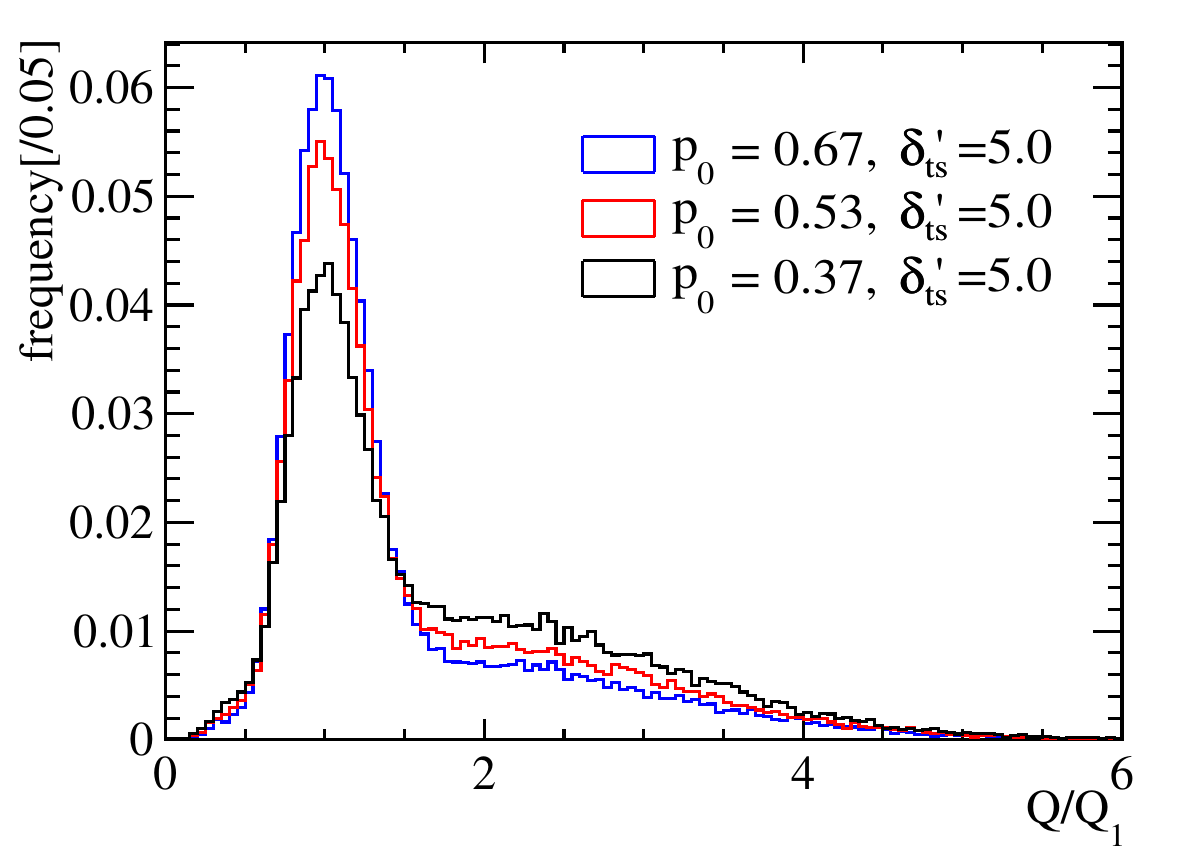}
        \caption{}
        \label{fig:p}
    \end{subfigure}
    \hfill
    \begin{subfigure}{0.47\textwidth}
        \centering
        \includegraphics[width=\linewidth]{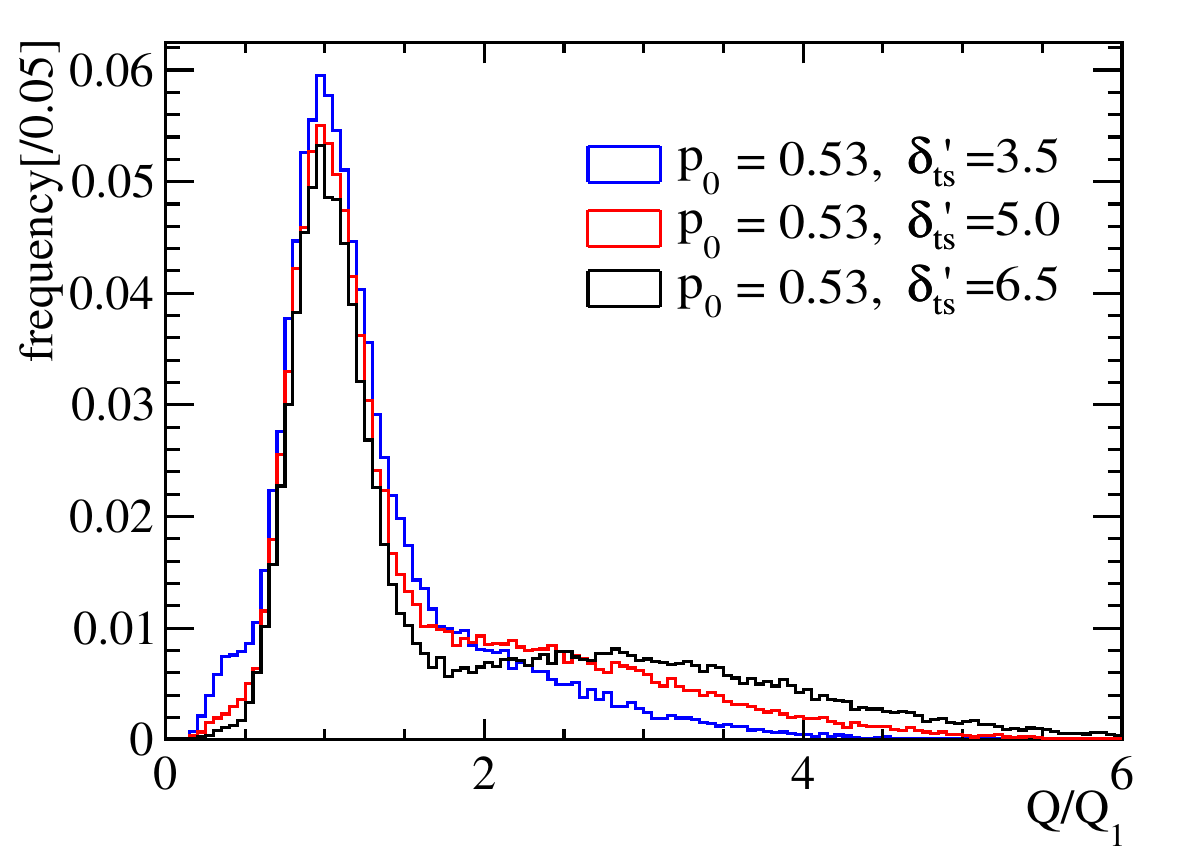}
        \caption{}
        \label{fig:ts}
    \end{subfigure}
    \caption{$\delta_{\mathrm{ts}}'$ and $p_0$ influence the shape of SER charge spectrum from MC.
        As $\delta_{\mathrm{ts}}'$ increases, the region of the tail becomes more prolonged.
        As $p_0$ increases, the height of the principal peak region increases, and the tail becomes narrower.
    }
    \label{fig:tsp}
\end{figure}

It is evident from Eq.~(\ref{eq:1}) and (\ref{eq:ts_all}) that $\delta_{\mathrm{ts}}'$ and \(p_0\)
significantly impact the SER charge distribution, demonstrated in Fig.~\ref{fig:tsp}.  We use
the MCP-PMT test data by Aiqiang Zhang~et~al.~\cite{Zhang:2023ued} to determine the two parameters.

Between each pair of predicted and measured charge distributions, we perform a chi-square test.
These two histograms are divided into $r$ bins using the same binning method.
The entries in the \(i\)-th bin are $n_{\mathrm{i}}$ and $m_{\mathrm{i}}$, adding up to
$N = \sum_{{\mathrm{i}}=1}^{r}n_{\mathrm{i}}$ and $M = \sum_{{\mathrm{i}}=1}^{r}m_{\mathrm{i}}$.
The chi-square test indicates the similarity between two histograms~\cite{2006Comparison},
\begin{equation}
    \label{eq:chi}
    \chi^2_{r-1}=\sum_{{\mathrm{i}}=1}^r \frac{\left(n_{\mathrm{i}}-N \hat{k}_{\mathrm{i}}\right)^2}{N \hat{k}_{\mathrm{i}}}+\sum_{{\mathrm{i}}=1}^r
    \frac{\left(m_{\mathrm{i}}-M \hat{k}_{\mathrm{i}}\right)^2}{M \hat{k}_{\mathrm{i}}}=\frac{1}{M N} \sum_{{\mathrm{i}}=1}^r
    \frac{\left(M n_{\mathrm{i}}-N m_{\mathrm{i}}\right)^2}{n_{\mathrm{i}}+m_{\mathrm{i}}}
\end{equation}
where \(\hat{k}_{\mathrm{i}}=\frac{n_{\mathrm{i}}+m_{\mathrm{i}}}{N+M}\).

\begin{figure}[!htbp]
    \centering
    \begin{subfigure}{0.47\textwidth}
        \centering
        \includegraphics[width=\linewidth]{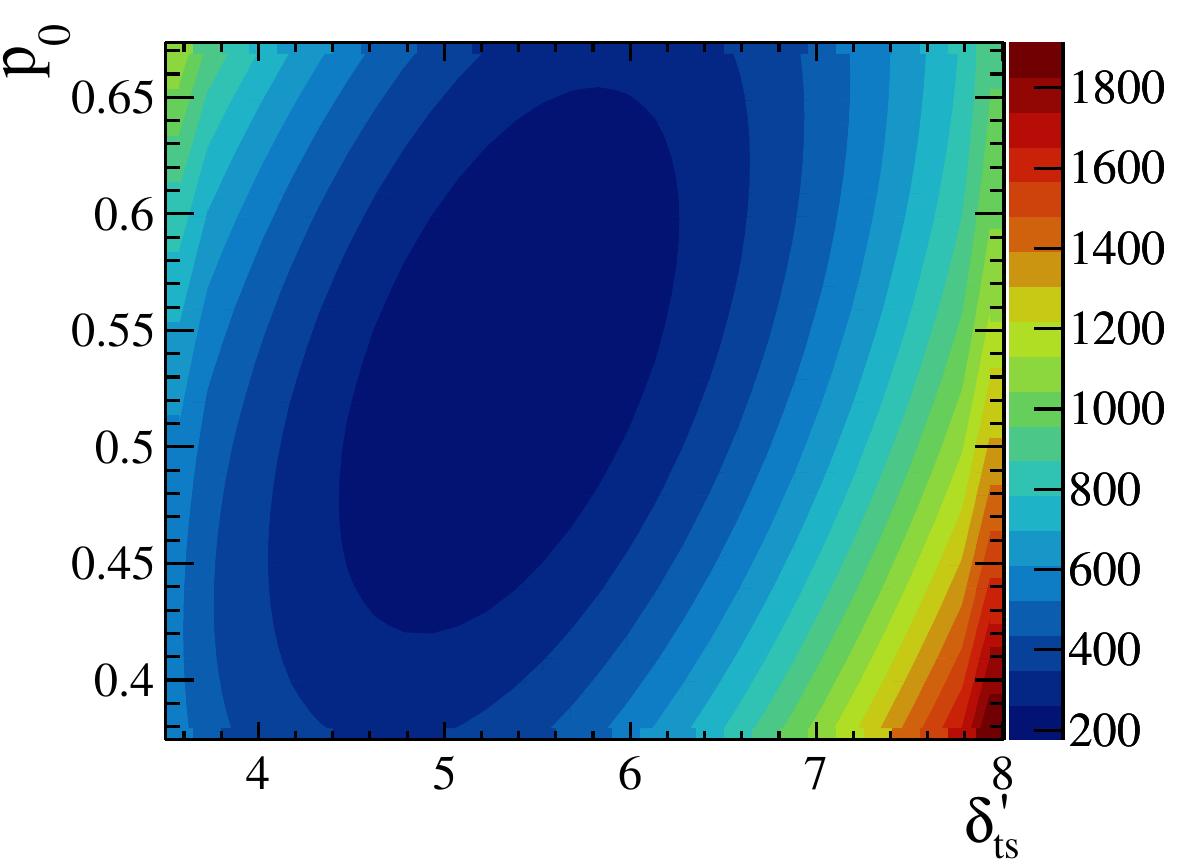}
        \caption{}
        \label{fig:cour}
    \end{subfigure}
    \hfill
    \begin{subfigure}{0.47\textwidth}
        \centering
        \includegraphics[width=\linewidth]{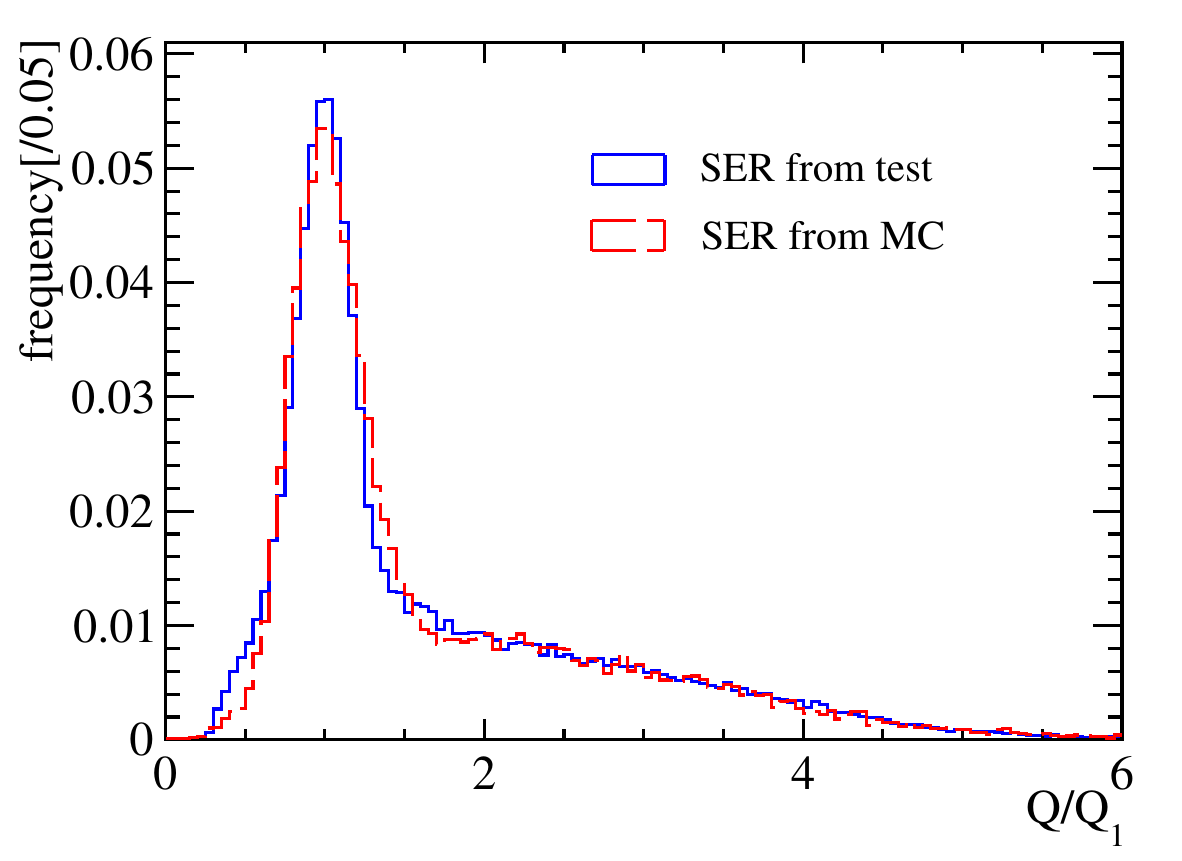}
        \caption{}
        \label{fig:hist}
    \end{subfigure}
    \caption{The plot~(\subref{fig:cour}) is the contour plot of the chi-square test, with $p_0$ and $\delta_{\mathrm{ts}}'$ as parameters
        and the chi-square values as the height.
        The plot~(\subref{fig:hist}) is an example of the MC histogram~(the red line) and the histogram from test~(the blue line).
    }
    \label{fig:chi}
\end{figure}

The \(\chi^2_{r-1}\) are scanned in the $(p_0,\delta_{\mathrm{ts}}')$ grid, with an example in Fig.~\ref{fig:cour}.
We use a linear model~\cite{Gelman_Hill_2006} to smooth the approximate parabolic relationship
between the \(\chi^2_{r-1}\) and $(p_0, \delta_{\mathrm{ts}}')$,
then extract the $(\hat{p}_0, \hat{\delta}_{\mathrm{ts}}')$ that minimizes \(\chi^2_{r-1}\) with
intervals at \SI{68.3}{\percent} confidence levels~\cite{cowan1997statistical}.

\begin{figure}[!htbp]
    \centering
    \includegraphics[width=0.6\textwidth]{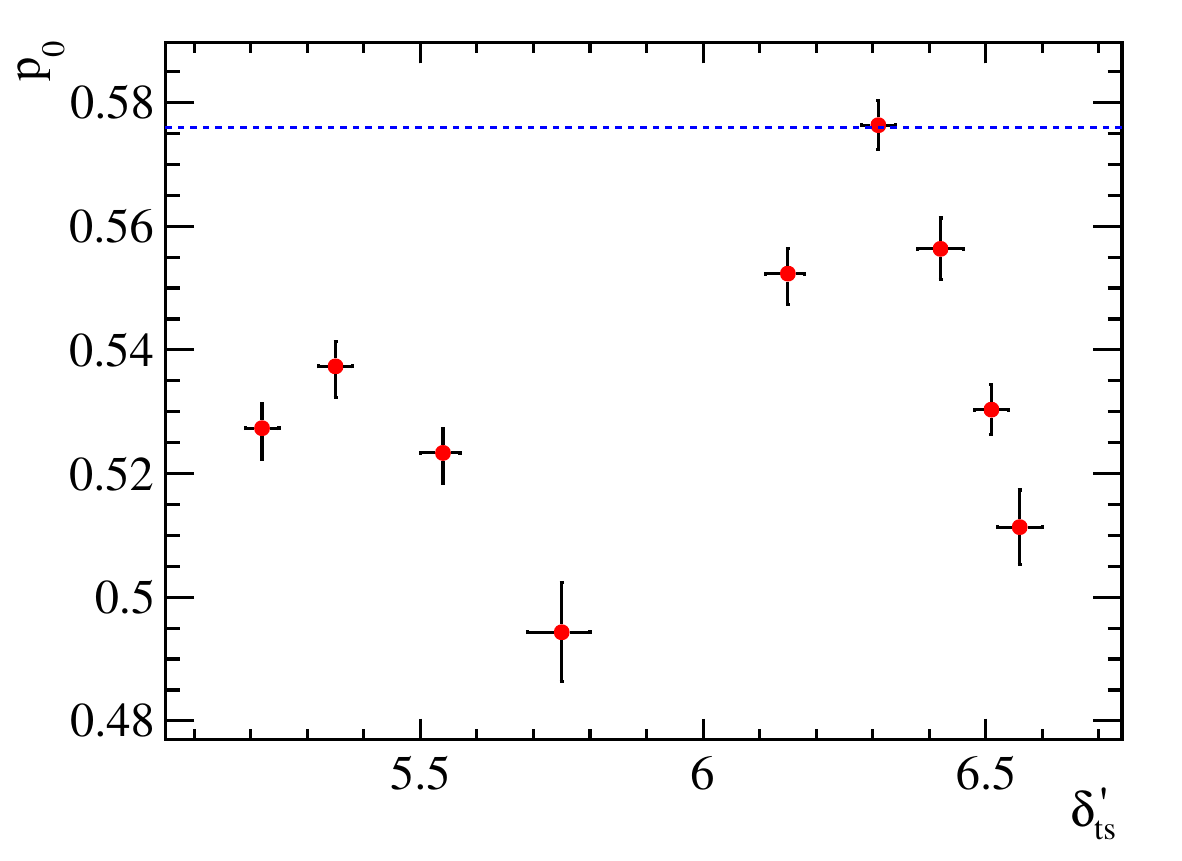}
    \caption{When convolving with 9 MCP-PMTs,
        the distribution of $\delta_{\mathrm{ts}}'$ and $p_0$ at the minimum chi-square
        occurs. The blue dashed line shows the expected $\hat{p}_0$ estimated from~\cite{chen2018photoelectron}.}
    \label{fig:true_p}
\end{figure}

The $\hat{\delta}_{\mathrm{ts}}'$ and $\hat{p}_0$ scatter plot
of 9 MCP-PMTs in Fig~\ref{fig:true_p} does not indicate a strong correlation.
They are determined by independent manufacturing stages.
On average, $\delta_{\mathrm{ts}}'$ is 5.979 and $p_0$ is 0.5341. 
The PEs of the channel, back-scattered and rediffused surface modes account for \SI{53.41}{\percent}.
They constitute the main peak. Each of the rest hits the surface
to induce 5.979 true-secondary electrons on average.

To compare our measurement to previous studies, we convert \(\delta_\text{ts}'\)
to the SEY \(\delta\)
\begin{equation}
    \label{eq:2}
    \delta = \delta_\text{bs} + \delta_\text{rd} + (1-\delta_\text{bs} - \delta_\text{rd}) \delta_\text{ts}'
\end{equation}
and the fraction of main peak \(p_0\) to that of the channel mode \(p\) by Eq.~\eqref{eq:p0}.
Weiwei Cao~et~al.\cite{cao_secondary_2021} measured the SEY of \ce{Al2O3}-\ce{MgO} double-layered film
to be 4--5. Lin Chen~et~al.~\cite{2016Optimization} pointed out
that there is an electrostatic lens effect at the MCP channel entrances,
resulting in the ratio of the PEs entering the MCP channels
being smaller than the open-area fraction.
When PEs come from the top of the MCP-PMT,
the proportion of the PEs directly entering the MCP channels is around \SI{60}{\percent} when the MCP open area fraction is \SI{74.9}{\percent}.
Ping Chen~et~al.~\cite{chen2018photoelectron} indicated that
the proportion is around \SI{55}{\percent} when the open area fraction is \SI{65}{\percent}.
The MCPs used in our tested MCP-PMTs have a pore diameter of \SI{12}{\mu\meter}, a spacing of \SI{14}{\mu\meter} between the pores,
and an open-area ratio of \SI{66.6}{\percent},
so the expected $\hat{p}_0\coloneqq \frac{55\%}{1-(1-55\%)(\delta_{\mathrm{rd}}+\delta_{\mathrm{bs}})}\approx 57.6\%$ for $\delta_{\mathrm{rd}}+\delta_{\mathrm{bs}}=0.1$.
\begin{figure}[!htbp]
    \centering
    \begin{subfigure}{0.48\textwidth}
        \centering
        \includegraphics[width=\linewidth]{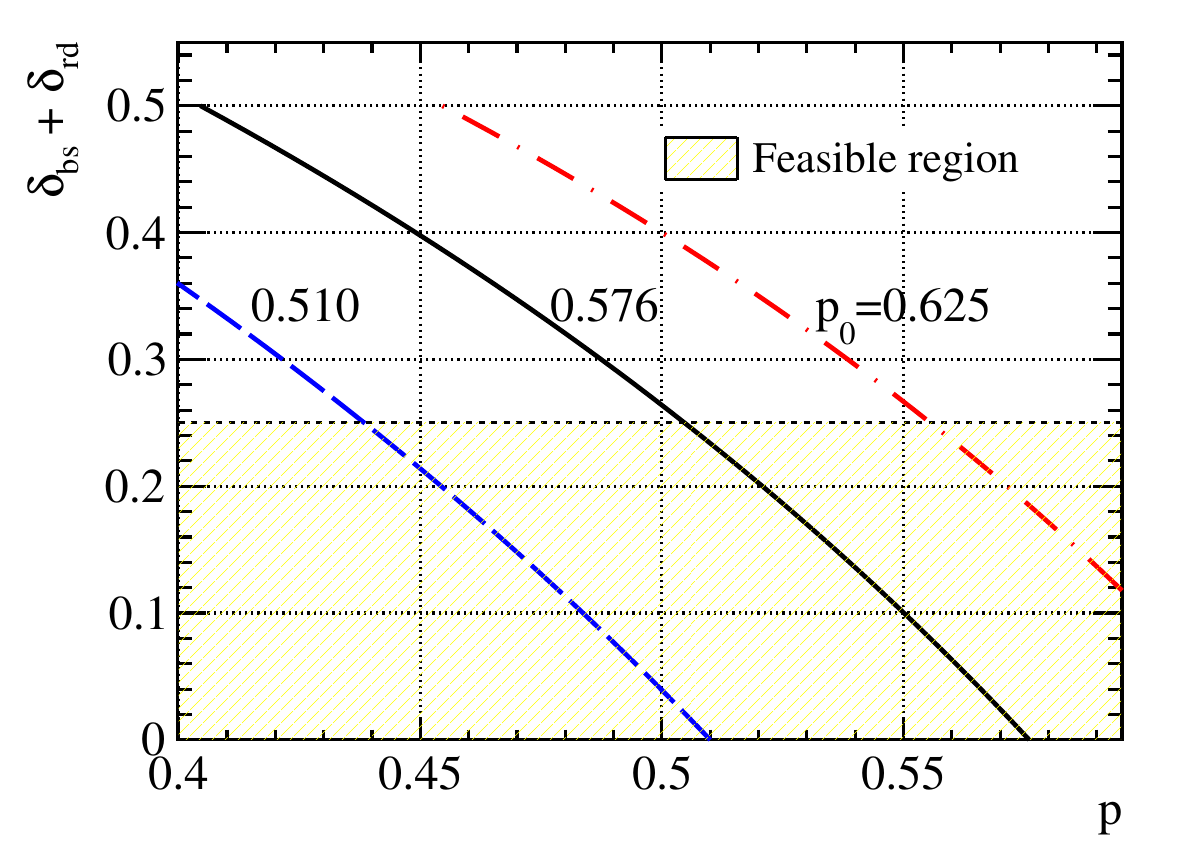}
        \caption{}
        \label{fig:pp0}
    \end{subfigure}
    \hfill
    \begin{subfigure}{0.48\textwidth}
        \centering
        \includegraphics[width=\linewidth]{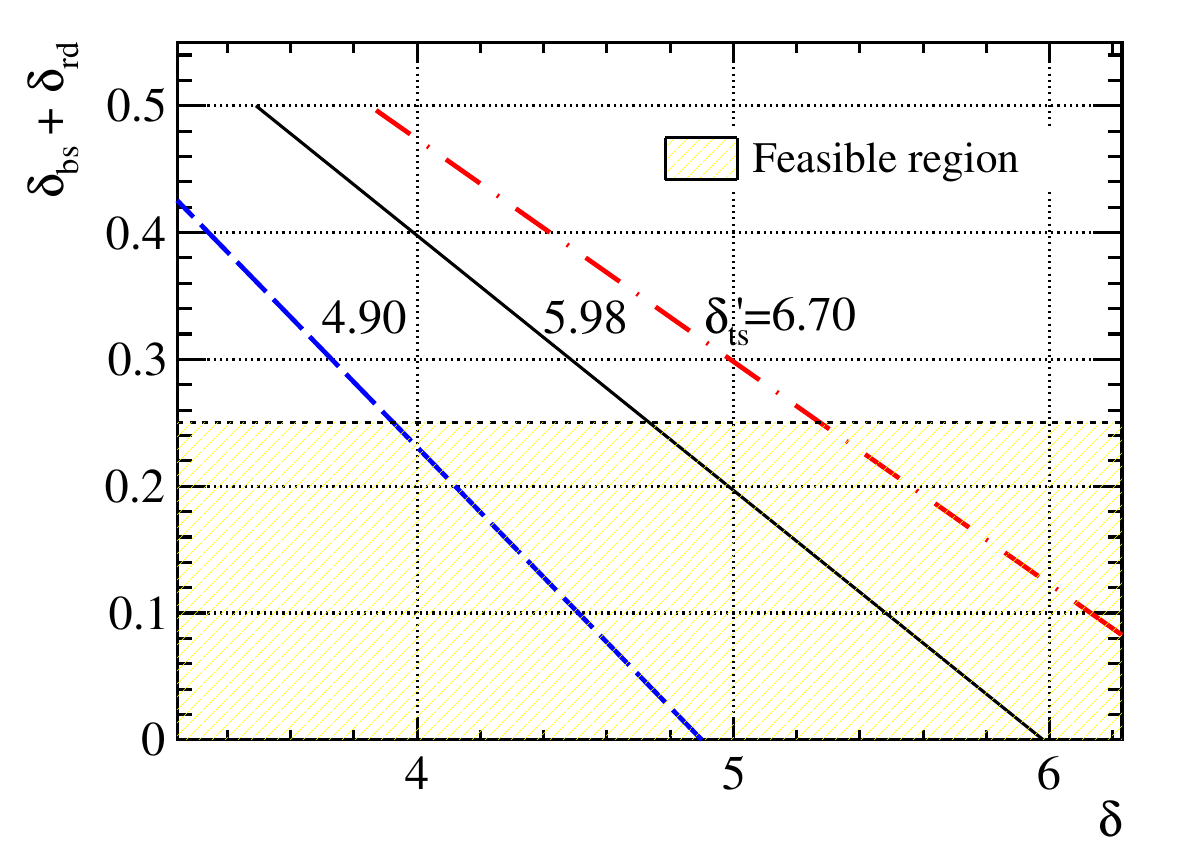}
        \caption{}
        \label{fig:tsts}
    \end{subfigure}
    \caption{Relations of \(\delta_\text{bs} + \delta_\text{rd}\) against the SEY \(\delta\) and the fraction of channel mode \(p\).
        The feasible region shows the consistency of our measurement to the literature.}
    \label{fig:pdelta}
\end{figure}

In Fig.~\ref{fig:pdelta}, with the typical values of \(\delta=5\) and \(p=0.55\),
our measurement is consistent with an assumption that \(\delta_\text{bs} + \delta_\text{rd} < 0.25\).
The small contribution of the back-scattered and rediffused electrons in SEE is pointed out by Beck~\cite{beck_physical_1966} to be especially true for insulators with high SEY.

\section{Discussion}\label{sec:discussion}
\subsection{Model Simplification with Tweedie}\label{sec:model}
\begin{figure}[!htbp]
    \centering
    \begin{subfigure}{0.47\textwidth}
        \centering
        \includegraphics[width=\linewidth]{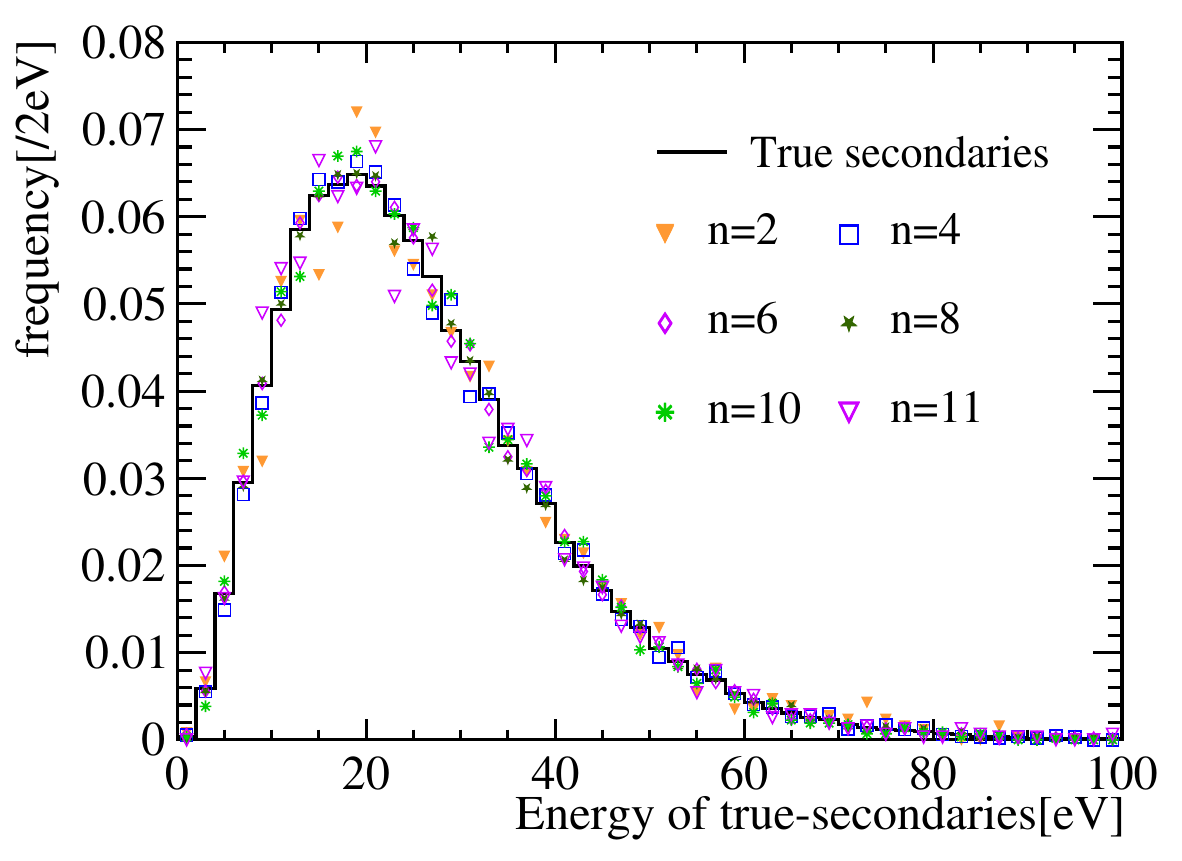}
        \caption{}
        \label{fig:single_pe}
    \end{subfigure}
    \hfill
    \begin{subfigure}{0.47\textwidth}
        \centering
        \includegraphics[width=\linewidth]{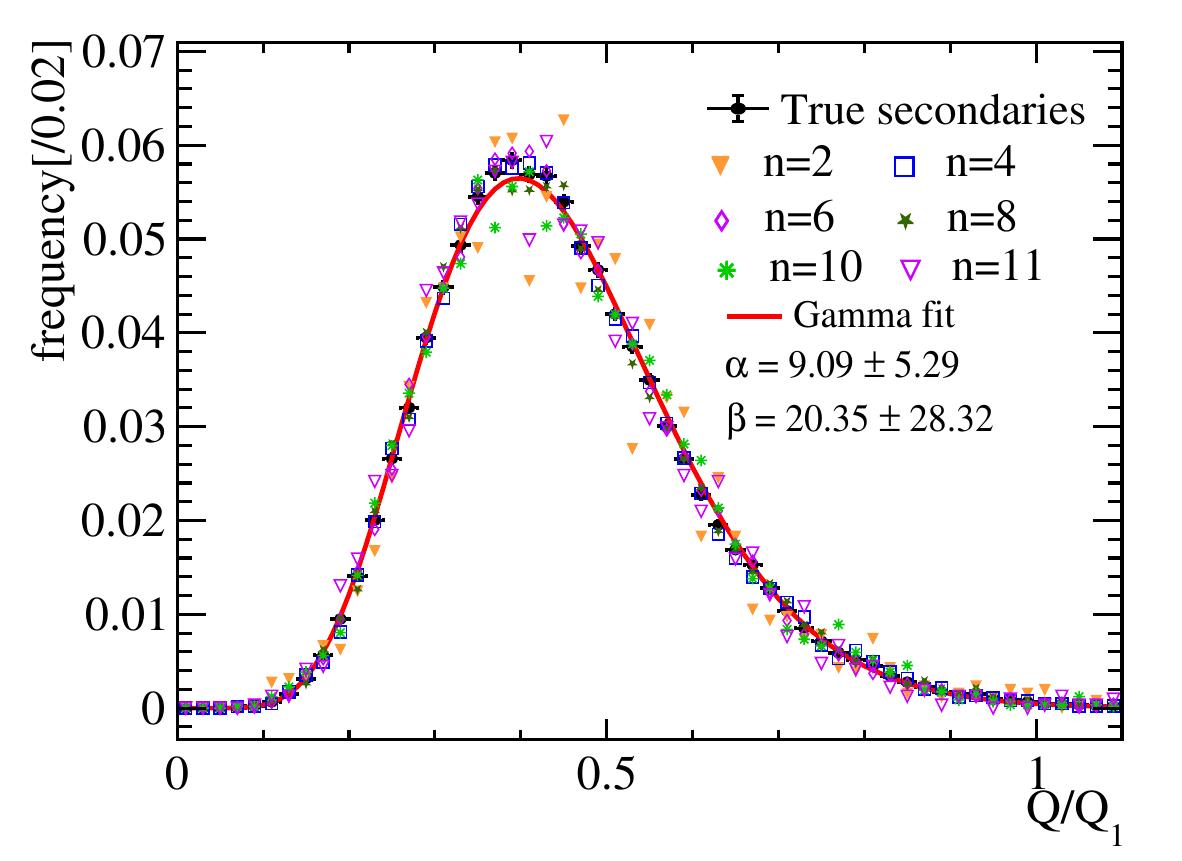}
        \caption{}
        \label{fig:single_fit}
    \end{subfigure}
    \caption{The energy distribution of and the charge response distribution of MCP to a single true-secondary electron when $n$ is different.
        (\subref{fig:single_pe}) all the energies of the true secondaries follow the same distribution,
        although $n$ is different.
        (\subref{fig:single_fit}) the charge response of MCP to a single true-secondary electron is identical, 
        and the fitting of the Gamma distribution~\(\varGamma[\alpha',\beta']\)
        achieves sufficient goodness.}
    \label{fig:singlepe}
\end{figure}

In our calculation, the distribution of the MCP charge response to the true-secondary electrons $\varGamma(\alpha_{\mathrm{i}},\beta_{\mathrm{i}})$ is determined by their energies $E_{\mathrm{i}}$,
which satisfy $\sum_{\mathrm{i}}^{\mathrm{n}}E_{\mathrm{i}}<E_0$.
The incident energy \(E_0\) of the PEs is \SI{650}{eV},
which is more than ten times the energies of the true secondaries.
Because $n$ follows the Poisson distribution with an expectation between 5 and 6.5,
the probability of $n$ exceeding 10 is negligible.
Thus the effect of $n$ on $E_{\mathrm{i}}$ can be ignored
and the energy \(E_{\mathrm{i}}\) is independently and identically distributed, as demonstrated in Fig.~\ref{fig:single_pe}.
The charge response of MCP to a single true-secondary electron in turn can be treated identically as shown in Fig.~\ref{fig:single_fit}.
Furthermore, a single Gamma distribution \(\varGamma[\alpha',\beta']\)
is flexible enough to describe the continuous mixture of 
\(\int dE_{\mathrm{i}}\frac{1}{\delta_{\mathrm{ts}}}\frac{d\delta_{\mathrm{ts}}}{dE_{\mathrm{i}}}\varGamma[\alpha(E_{\mathrm{i}}),\beta(E_{\mathrm{i}})]\).

When we use such a single \(\varGamma(\alpha', \beta')\) in Eq.\eqref{eq:ts_all},
the resulting Poisson-Gamma compound is a special case of the Tweedie distribution $\mathrm{Tw}_{\xi}(\alpha,\beta)$
for $1<\xi<2$~\cite{1991Tweedie}.
\begin{equation}
  \arraycolsep=1.4pt
  \label{eq:ts_gamma}
  \left.\begin{array}{rl}
          Q_{\text{ts}} &= \sum_{{\mathrm{i}}=1}^{n} Q_{\mathrm{i}}      \\
          n &\sim \mathrm{\pi}(\delta_{\mathrm{ts}}')\\
          Q_{\mathrm{i}} &\sim \varGamma(\alpha', \beta')
        \end{array}\right\} \implies
      Q_{\text{ts}} \sim \mathrm{Tw}_{\xi}(\alpha',\beta')
\end{equation}
A phenomenological joint fit of the \(f_\mathrm{ch}\) Gamma and \(f_\mathrm{ts}\) Tweedie
mixture with Eq.~\eqref{eq:1} and \eqref{eq:ts_gamma} is sufficient to calibrate the SER charge spectrum and measure \(p_0\) and \(\delta_\text{ts}'\).
To put it differently, the voltage division experiment~(Sec.~\ref{sec:gain}) relations \(\mu(E_\mathrm{i})\)/\(\sigma(E_\mathrm{i})\)
and the Furman model provides the understanding of the jumbo charges and the justification of the phenomenological Gamma-Tweedie
mixture, but is less practically useful in PMT calibrations.

The number of parameters, 2 for \(f_\mathrm{ch}\) Gamma and 3 for \(f_\mathrm{ts}\) Tweedie, hinders convergence
unless we aided it with physical constraints.
Typically $\frac{\alpha'}{\beta'}\approx 0.45Q_1$ and \(\sqrt{\frac{\alpha'}{\beta'^2}}\approx 0.15Q_1\). 
It is practical to bound them in $[0.3,0.7]Q_1$ and $[0.05,0.3]~Q_1$ when the incident energy $E_0$ is significantly greater than $E_{\mathrm{i}}$.
We also have checked the chi-square results in the Gamma-Tweedie fitting, 
which gives good $\chi^2/\mathrm{ndf}<10$. 

\subsection{Transit Time Characteristics}\label{sec:time}
Unlike the channel mode electrons, secondaries from the surface move away from the MCP
before being drawn back by the electric field. The elastically back-scattered electron has a typical round trip
time of \SI{40}{ns}~\cite{Zhang:2023ued}. The true secondaries have smaller kinetic energies.
Our MCP-PMT uses Ping Chen~et~al.~\cite{chen2018photoelectron}'s design of an extra focusing
electrode in front of the MCP to switchly collect low-energy electrons in less than \SI{1}{ns}.
To separate the surface true secondaries from the channel electrons, high-precision timing
electronics and accurate calibration of optical systems are necessary.
We are planning a sub-nanosecond transit time measurement to verify the delay and the proportion of the true secondaries.

\subsection{Offsetting the Harm of Jumbo Charges}
The jumbo charge is a negative by-product of the high collection efficiency
that reduces the charge resolution~\cite{Zhang:2023ued}.
Primarily, there are two ways to tackle the issue.
The first approach addresses the generation of the jumbo charges by
designing MCPs with larger open areas
to decrease the proportion of the PEs hitting the MCP surface.
The second method entails an extension of fast stochastic matching pursuit~\cite{Wang_2024} for this type of MCP-PMT.
We are developing charge calibration methods specifically for this type of MCP-PMT
to utilize its performance fully.

\subsection{Scope of Our Model}\label{sec:scope}
Our Gamma-Tweedie SER charge spectrum model for the MCP-PMT can be extended to the Dynode-PMT.
In Sec.~\ref{sec:see} and Fig~\ref{fig:circuit}, we introduced the surface as an extra stage of multiplication
for the jumbo charges. When the Dynode-PMT generates a pre-pulse, the electron is produced by the light hitting the first dynode,
and begins to be amplified from the second dynode resulting in a loss of multiplication, giving a charge similar
to a PE from the photocathode missing the first dynode. Such lack-of-multiplication can be
modeled the same way as the extra-multiplication by Gamma-Tweedie mixtures. If the Tweedie parameter \(\xi\)
is close to 2, it reduces to another Gamma distribution so that the mixture becomes dual Gamma.

\section{Conclusion}\label{sec:conclusion}
The 8-inch MCP-PMTs in this study have maximal CEs but exibit the jumbo charges in the SER charge spectrum.
We successfully find the nature behind it with the theory of SEE.
By employing a dual high-voltage circuit in the voltage-division experiment, 
the MCP gain for the electrons with varying energies can be measured.
The origin of the jumbo charges is that the PEs hit the input electrode of the first MCP,
and generate multiple true-secondary electrons entering the channels for amplification.

The calculation of the SER charge spectrum of the 8-inch high-CE MCP-PMT is achieved.
The yield of the true-secondary electrons from penetrating electrons is measured to be around 5.979,
making the first study on the phenomenon of SEE in pulse mode at a working PMT.
Based on the nature of the jumbo charges, we propose a new Gamma-Tweedie mixture model for the SER charge spectrum.

\section{Acknowledgments}
Special thanks to Ling Ren and North Night Vision Science \& Technology (Nanjing) Research Institute Co. Ltd. (NNVT)
for providing all the PMTs and the voltage dividers we used for this study.
Also, we would like to thank Yiqi Liu, Jiashen Tao, Chuang Xu, Xin Wang, and others for their assistance in the experiment,
as well as Yuyi Wang for help with data analysis. Their support and assistance have been crucial to this study.
This work is supported in part by the National Natural Science Foundation of China (12127808, 12335012),
National key research and development program of China~(Grant no. 2023YFC3107402),
and the Key Laboratory of Particle \& Radiation Imaging (Tsinghua University).

\bibliographystyle{unsrt}
\bibliography{ref}

\begin{thebibliography}{10}

\bibitem{1955Scintillation}
Ernst Breitenberger.
\newblock Scintillation spectrometer statistics.
\newblock {\em Progress in nuclear physics}, 4:56--94, 1955.

\bibitem{2016Optimization}
Lin Chen et~al.
\newblock Optimization of the electron collection efficiency of a large area
  {MCP-PMT} for the {JUNO} experiment.
\newblock {\em Nucl. Instrum. Meth. A}, 827:124--130, 2016.

\bibitem{1994Absolute}
E.H. Bellamy et~al.
\newblock Absolute calibration and monitoring of a spectrometric channel using
  a photomultiplier.
\newblock {\em Nucl. Instrum. Meth. A}, 339(3):468--476, 1994.

\bibitem{2016Secondary}
Shu~Xia Tao, Hong~Wah Chan, and Harry Van~der Graaf.
\newblock Secondary electron emission materials for transmission dynodes in
  novel photomultipliers: A review.
\newblock {\em Materials}, 9(12):1017, 2016.

\bibitem{2002Probabilistic}
MA~Furman and MTF Pivi.
\newblock Probabilistic model for the simulation of secondary electron
  emission.
\newblock {\em Physical review special topics-accelerators and beams},
  5(12):124404, 2002.

\bibitem{1938Secondary}
H~Bruining and JH~De~Boer.
\newblock {Secondary electron emission: Part I. Secondary electron emission of
  metals}.
\newblock {\em Physica}, 5(1):17--30, 1938.

\bibitem{1988Secondary}
Y~Ushio et~al.
\newblock Secondary electron emission studies on {MgO} films.
\newblock {\em Thin Solid Films}, 167(1-2):299--308, 1988.

\bibitem{2012Secondary}
Slade~J. Jokela et~al.
\newblock {Secondary Electron Yield of Emissive Materials for Large-Area
  Micro-Channel Plate Detectors: Surface Composition and Film Thickness
  Dependencies}.
\newblock {\em Physics Procedia}, 37:740--747, 2012.

\bibitem{OLANO2020103456}
L.~Olano and I.~Montero.
\newblock Energy spectra of secondary electrons in dielectric materials by
  charging analysis.
\newblock {\em Results in Physics}, 19:103456, 2020.

\bibitem{2012An}
Anil~U. Mane et~al.
\newblock {An Atomic Layer Deposition Method to Fabricate Economical and Robust
  Large Area Microchannel Plates for Photodetectors}.
\newblock {\em Physics Procedia}, 37:722--732, 2012.

\bibitem{2021Effects}
Lehui Guo et~al.
\newblock Effects of secondary electron emission yield properties on gain and
  timing performance of {ALD-coated MCP}.
\newblock {\em Nucl. Instrum. Meth. A}, 1005:165369, 2021.

\bibitem{Nathan1970TheED}
R~Nathan and CHB Mee.
\newblock {The energy distribution of photoelectrons from the K2CsSb
  photocathode}.
\newblock {\em physica status solidi (a)}, 2(1):67--72, 1970.

\bibitem{ZHU2020162002}
Yao Zhu et~al.
\newblock The mass production and batch test result of {20" MCP-PMTs}.
\newblock {\em Nucl. Instrum. Meth. A}, 952:162002, 2020.
\newblock 10th International Workshop on Ring Imaging Cherenkov Detectors (RICH
  2018).

\bibitem{Zhang:2023ued}
Aiqiang Zhang et~al.
\newblock {Performance evaluation of the 8-inch MCP-PMT for Jinping Neutrino
  Experiment}.
\newblock {\em Nucl. Instrum. Meth. A}, 1055:168506, 2023.

\bibitem{MATSUOKA2014148}
K.~Matsuoka.
\newblock {Development and production of the MCP-PMT for the Belle II TOP
  counter}.
\newblock {\em Nucl. Instrum. Meth. A}, 766:148--151, 2014.

\bibitem{KRAUSS2023168659}
S.~Krauss et~al.
\newblock {Performance of the most recent Microchannel-Plate PMTs for the PANDA
  DIRC detectors at FAIR}.
\newblock {\em Nucl. Instrum. Meth. A}, 1057:168659, 2023.

\bibitem{Cao2019UpgradingPT}
Zhen Cao, M.~J. Chen, H.~C. Li, and Z.~G. Yao.
\newblock {Upgrading Plan Towards Multi-messenger Observation with LHAASO}.
\newblock {\em EPJ Web of Conferences}, 2019.

\bibitem{N2006Lifetime}
N.~Kishimoto et~al.
\newblock Lifetime of {MCP-PMT}.
\newblock {\em Nucl. Instrum. Meth. A}, 564(1):204--211, 2006.

\bibitem{reviewer2}
D.R. Beaulieu et~al.
\newblock Nano-engineered ultra-high-gain microchannel plates.
\newblock {\em Nucl. Instrum. Meth. A}, 607(1):81--84, 2009.
\newblock Radiation Imaging Detectors 2008.

\bibitem{Lehmann:2022ret}
A~Lehmann et~al.
\newblock {Latest Technological Advances with MCP-PMTs}.
\newblock {\em J. Phys.: Conf. Ser.}, 2374(1):012128, November 2022.

\bibitem{cao_secondary_2021}
Weiwei Cao et~al.
\newblock Secondary electron emission characteristics of the {Al2O3}/{MgO}
  double-layer structure prepared by atomic layer deposition.
\newblock {\em Ceramics International}, 47(7):9866--9872, April 2021.

\bibitem{zzj2021Al}
Zhengjun ZHANG, Xiangbiao QIU, Fangjian QIAO, et~al.
\newblock {Effect of Al2O3/MgO Composite Layer on the Properties of
  Microchannel Plate}.
\newblock {\em Surface Technology}, 50(6):199--205, 2021.

\bibitem{JUNO:2022hlz}
Angel Abusleme et~al.
\newblock {Mass testing and characterization of 20-inch PMTs for JUNO}.
\newblock {\em The European Physical Journal C}, 82(12):1168, 2022.

\bibitem{reviewer1}
D.~A. Orlov, T.~Ruardij, S.~Duarte~Pinto, R.~Glazenborg, and E.~Kernen.
\newblock {High collection efficiency MCPs for photon counting detectors}.
\newblock {\em JINST}, 13(01):C01047, 2018.

\bibitem{2021Gain}
H.Q. Zhang et~al.
\newblock Gain and charge response of {20" MCP and dynode PMTs}.
\newblock {\em JINST}, 16(08):T08009, 2021.

\bibitem{2017MCP}
Yuzhen Yang et~al.
\newblock {MCP} performance improvement using alumina thin film.
\newblock {\em Nucl. Instrum. Meth. A}, 868:43--47, 2017.

\bibitem{branchandPoisson}
Harry~H Tan.
\newblock A statistical model of the photomultiplier gain process with
  applications to optical pulse detection.
\newblock {\em The Telecommunications and Data Acquisition Report}, April 1982.

\bibitem{Bartlett1963TheTO}
M.~S. Bartlett and Theodore~Edward Harris.
\newblock {\em {The Theory of Branching Processes}}.
\newblock Springer Berlin, 1963.

\bibitem{polya}
JR~Prescott.
\newblock A statistical model for photomultiplier single-electron statistics.
\newblock {\em Nuclear Instruments and Methods}, 39(1):173--179, 1966.

\bibitem{2012Calibration}
Leonidas Kalousis.
\newblock {\em Calibration of the {Double Chooz} detector and cosmic background
  studies}.
\newblock PhD thesis, University of Strasbourg, 2012.

\bibitem{2020A}
LN~Kalousis, JPAM De~Andr{\'e}, E~Baussan, and M~Dracos.
\newblock A fast numerical method for photomultiplier tube calibration.
\newblock {\em JINST}, 15(03):P03023, 2020.

\bibitem{bruining_physics_1954}
H.~Bruining.
\newblock {\em Physics and Applications of Secondary Electron Emission}.
\newblock Pergamon Press, 1954.

\bibitem{baroody1950theory}
EM~Baroody.
\newblock A theory of secondary electron emission from metals.
\newblock {\em Physical Review}, 78(6):780, 1950.

\bibitem{dekker1952theory}
AJ~Dekker and A~Van~der Ziel.
\newblock Theory of the production of secondary electrons in solids.
\newblock {\em Physical Review}, 86(5):755, 1952.

\bibitem{wolff1954theory}
P.~A. Wolff.
\newblock Theory of secondary electron cascade in metals.
\newblock {\em Phys. Rev.}, 95:56--66, Jul 1954.

\bibitem{Kanaya_1978}
K.~Kanaya, S.~Ono, and F.~Ishigaki.
\newblock Secondary electron emission from insulators.
\newblock {\em J. Phys. D: Appl. Phys.}, 11(17):2425, dec 1978.

\bibitem{vaughan}
J.R.M. Vaughan.
\newblock A new formula for secondary emission yield.
\newblock {\em IEEE Transactions on Electron Devices}, 36(9):1963--1967, 1989.

\bibitem{Luo:2023jdf}
Feng-Jiao Luo et~al.
\newblock {Design $\&$ Optimization of the HV divider for JUNO 20-inch PMT}.
\newblock {\em arXiv preprint arXiv:2307.10544}, 2023.

\bibitem{teledynelecroy}
Nelson and Rick.
\newblock {High-Definition Oscilloscopes Optimize Vertical Resolution}.
\newblock {\em EE-Evaluation Engineering Online}, 2016.

\bibitem{Xu_2022}
D.C. Xu et~al.
\newblock Towards the ultimate {PMT} waveform analysis for neutrino and dark
  matter experiments.
\newblock {\em JINST}, 17(06):P06040, jun 2022.

\bibitem{1951Various}
Von Neumann.
\newblock Various techniques used in connection with random digits.
\newblock {\em Notes by GE Forsythe}, pages 36--38, 1951.

\bibitem{2006Comparison}
N.~D Gagunashvili.
\newblock Comparison of weighted and unweighted histograms.
\newblock {\em Statistics}, pages 43--44, 2012.

\bibitem{Gelman_Hill_2006}
Andrew Gelman and Jennifer Hill.
\newblock {\em {Data Analysis Using Regression and Multilevel/Hierarchical
  Models}}.
\newblock Analytical Methods for Social Research. Cambridge University Press,
  2006.

\bibitem{cowan1997statistical}
Glen Cowan.
\newblock {\em {Statistical Data Analysis}}.
\newblock Oxford University Press, Oxford: New York, illustrated edition
  edition, 1997.

\bibitem{chen2018photoelectron}
Ping Chen et~al.
\newblock Photoelectron backscattering in the microchannel plate
  photomultiplier tube.
\newblock {\em Nucl. Instrum. Meth. A}, 912:112--114, 2018.

\bibitem{beck_physical_1966}
A.~H. Beck.
\newblock {\em Physical {Electronics}: {Handbook} of {Vacuum} {Physics}}.
\newblock Elsevier, 1966.

\bibitem{1991Tweedie}
B.~Jorgensen.
\newblock {\em {The Theory of Dispersion Models}}.
\newblock Taylor \& Francis, 1997.

\bibitem{Wang_2024}
Yuyi Wang et~al.
\newblock {{The Fast Stochastic Matching Pursuit for Neutrino and Dark Matter
  Experiments}}, 3 2024.

\end{thebibliography}
\end{document}